\documentclass[aps,floatfix,showpacs]{revtex4}
\usepackage{natbib}
\usepackage{dcolumn}
\usepackage{graphicx}

\newcommand{\cals}{\zeta_s}
\newcommand{\calr}{\zeta_r}

\newcommand{\mnras}{Mon.~Not.~R.~Astron.~Soc.}

\newcommand{\aj}{Astron.~J.}
\newcommand{\physrep}{Phys.~Rep.}
\newcommand{\araa}{Ann.~Rev.~Astron.~Astrophys.}
\newcommand{\apjs}{Astrophys.~J.~Supp.}

\begin{document}
\title{Remarks on the formulation of the cosmological constant/dark energy problems} 
\author{Mustapha Ishak$^{1}$\cite{email},}

\affiliation{
$^1$ Department of Physics, The University of Texas at Dallas, Richardson, TX 75083, USA}
\date{\today}
\begin{abstract}
Associated with the cosmic acceleration are the old and new cosmological constant 
problems, recently put into the more general context of the dark energy problem. 
In broad terms, the old problem is related to an unexpected order of magnitude of 
this component while the new problem is related to this magnitude being of the 
same order of the matter energy density during the present epoch of cosmic 
evolution. Current plans to measure the equation of state or density parameters 
certainly constitute an
important approach; however, as we discuss, this approach is faced with serious
feasibility challenges and is limited in the type of conclusive answers it could
provide. Therefore, is it really too early to seek actively for new tests and
approaches to these problems? In view of the difficulty of this endeavor, we argue
in this work that a good place to start is by questioning some of the assumptions
underlying the formulation of these problems and finding new ways to put this
questioning to the test. 
First, we calculate how much fine tuning the cosmic coincidence problem represents.
Next, we discuss the potential of some cosmological probes such as weak
gravitational lensing to identify novel tests to probe dark energy questions and
assumptions and provide an example of consistency tests. Then, motivated by some 
theorems in General Relativity, we discuss if the full identification of the 
cosmological constant with vacuum energy is unquestionable. 
We discuss some implications of the simplest solution for the principles of General
Relativity. Also, we point out the relevance of experiments at the interface of
astrophysics and quantum field theory, such as the Casimir effect in gravitational
and cosmological contexts. We conclude that challenging some of the assumptions
underlying the cosmological constant problems and putting them to
the test may prove useful and necessary to make progress on these questions.
\end{abstract}
\maketitle
\section{Introduction}
\label{sec:intro}
Several different types of astrophysical observations, {\it e.g.}
\cite{supernovae,cmb,Spergel2003,galaxy,clusters,hubble,isw}, have established 
the evidence that the expansion of the universe entered a phase of 
acceleration. Associated with this acceleration is a cosmological 
constant, or another dark energy component, that contributes
significantly to the total energy density of the universe.
Cosmic acceleration and dark energy constitute one of the most 
important and challenging of current problems in cosmology and other areas 
of physics. There are many comprehensive reviews of the cosmological 
constant or dark energy, including the observational evidence 
for it and the problems associated with it, and we refer the reader to 
some of them \cite{review1,review2,review3,review4,review5,review6,review7,review8}.

Three questions related to the cosmic acceleration are 
encountered in the published literature, two of which are found in 
different formulations or expressions.  These are:

i) What is causing the cosmic acceleration? Is it a cosmological 
constant or a dynamical dark energy component \cite{quintessence}? 
Is this associated with the stress-energy momentum sector or 
with the curvature sector of the Einstein field equations (EFE)? 
Is this an indication of new physics at cosmological scales? 

ii) The old cosmological constant problem: 
If the acceleration is caused by vacuum energy, then why is the value
measured from astrophysics so small compared to values obtained from 
quantum field theory calculations (this is the puzzling smallness 
formulation, see for example \cite{review5})? 
Another formulation is: Why do all the contributions to the effective cosmological
constant term cancel each other up to a very large number of decimal
places (this is the fantastic cancelation formulation, see for example 
\cite{review1})? We contrast the two formulations in the next section.

iii) The new cosmological constant problem: 
Why is the acceleration happening during the present epoch
of the cosmic evolution? (Any earlier would have prevented 
structures from forming in the universe.) 
This is also formulated as the cosmic coincidence:
i.e. why is the dark energy 
density of the same order of magnitude as the matter 
density during the present time?

In this paper, we argue that questioning the formulation of these 
problems and challanging the underlying assumptions may proove 
useful and necessary in order to make progress on the questions.
We start by discussing the cosmic coincidence problem
and then make some clarifications using some fraction calculations. Then, we 
calculate some prospective constraints on the cosmological parameters 
of dark energy/cosmological constant and discuss the inferred possible 
answers. Constraining the equation of state parameters 
is certainly an important approach to persue, however, as we show, 
the level of precision required is very challanging, and yet the 
approach is limited in the kind of answers it could provide. 
This points out the need for new approaches to dark energy/cosmological constant 
problems. In particular, questioning the assumptions underlying these 
questions may be found useful in order to look for new types of tests or 
approaches to these problems. We provide an example where cosmological 
probes of the cosmic expansion and the growth rate of large-scale struture-formation 
can be used in order to distinguish between cosmic acceleration due to dark energy and 
cosmic acceleration due to some modification to General Relativity at cosmological 
scales. Next, we question the full identification of the cosmological constant with 
vacuum energy in light of some theorems on the most general curvature tensor in the 
Einstein Field equations. Then, we discuss how a simple intrinsic constant curvature 
of spacetime would fit within important interpretations of General Relativity's principles. 
Finally, we point out the relevance of experiments at the interface of particle physics 
and astrophysics, such as the Casimir effect in astrophysical and cosmological contexts. 
A discussion and conclusion are provided in the last section.
\section{Preliminaries}
\label{sec:preliminaries}
\subsection{Notation}
We recall here only some preliminary equations 
and definitions necessary for the clarity
of the paper.  We refer the reader to 
review papers, see e.g. \cite{review1,review2,review3,review4,review5,review6,review7} and
text books, see e.g. \cite{book1,book2,book3,book4}. The Einstein Field Equations (EFE)
with a cosmological constant $\Lambda$ read
\begin{equation}
\label{eq:efe}
G_{\alpha \beta}+\Lambda g_{\alpha \beta}=\kappa T_{\alpha \beta}
\end{equation}
where $\kappa\equiv 8\pi G$ and 
\begin{equation}
\label{eq:einstein}
G_{\alpha \beta}\equiv R_{\alpha \beta}-\frac{1}{2}g_{\alpha \beta} R
\end{equation}
is the Einstein tensor, $R_{\alpha \beta}$, $R$ and $g_{\alpha \beta}$
are the Ricci tensor, the Ricci curvature scalar and the metric tensor 
respectively. For a perfect fluid, the energy-momentum tensor is given by
\begin{equation}
T_{\alpha \beta}=(\rho+p)u_{\alpha}u_{\beta}+p g_{\alpha \beta}
\end{equation} 
where $u^{\alpha}$ is the four velocity vector and $\rho$ and $p$ are the 
energy-density and isotropic pressure relative to $u^{\alpha}$. 
With global isotropy, when $T_{\alpha \beta}=0$, the EFE (\ref{eq:efe}) 
admit de Sitter space (for $\Lambda>0$) as a solution, for which a line element is
given by (\ref{eq:desitter}) in the appendix.

Motivated by current observations, e.g. 
\cite{supernovae,cmb,Spergel2003,galaxy,clusters,hubble,isw}, and 
for the sake of simplicity, let us consider 
 the concordance spatially flat universe 
with a positive cosmological constant $\Lambda$.
The EFE (\ref{eq:efe}) with a dust source 
 can be solved 
explicitly to give 
\begin{equation}
\label{eq:scalefactor}
a(t)=\Big{(}\frac{3C}{\Lambda}\Big{)}^{1/3}\Big{[}
\sinh(\frac{\sqrt{3\Lambda}}{2}t)\Big{]}^{2/3}
\end{equation}
where $C\equiv 8\pi\rho a^3/3$ is a constant; and the 
 spacetime line element is given by 
\begin{equation}
\label{eq:frwdesitter}
ds^2=-dt^2+a^2(t)
(dr^2+r^2 d\Omega^2).
\end{equation}
where $d\Omega^2=(d\theta^2+\sin^2 \theta d\phi^2)$.
At early stages, the universe is matter dominated and 
$a(t)=(\frac{9C}{4}t^2)^{1/3}$. At late stages, the universe 
is dominated by the cosmological constant and enters a de Sitter 
phase with $a(t)=(\frac{3C}{4\Lambda})^{1/3}\exp{\sqrt{\Lambda/3}}t$.
We plot the curvature evolution and profile of these spacetimes in the appendix.

Now, to introduce the vacuum energy density, let us consider a 
scalar field with Lagrangian density
\begin{equation}
\mathcal{L}_{\phi}=-\frac{1}{2}g^{\alpha \beta}\partial_{\alpha}\phi\partial_{\beta}\phi
- V(\phi)
\end{equation}
 and energy momentum tensor
\begin{equation}
\label{eq:vacT1}
T^{\phi}_{\alpha \beta}=\partial_{\alpha}\phi\partial_{\beta}\phi
-\frac{1}{2}(g^{\delta
  \gamma}\partial_{\delta}\phi\partial_{\gamma}\phi)g_{\alpha 
  \beta}   
- V(\phi)g_{\alpha
  \beta} 
\end{equation}
The lowest energy density of the field configuration is when the 
kinetic (or gradient) term vanishes and the potential is at the
minimum $V(\phi_{min})$. The energy momentum (\ref{eq:vacT1}) reduces to 
\begin{equation}
\label{eq:vacT2}
T^{vac}_{\alpha \beta}=- V(\phi_{min})g_{\alpha
  \beta}=-\rho_{vac}g_{\alpha \beta} 
\end{equation}
This form of $T^{vac}_{\alpha \beta}$ is also the only
Lorentz-invariant form for the vacuum energy. 

\subsection{The old cosmological constant problem}
\label{sec:old_problem}
The common identification of the cosmological constant with
the vacuum energy density is based on the mathematical equivalence 
of the vacuum energy momentum tensor (\ref{eq:vacT2}) and the 
cosmological constant term on the LHS of the EFE (\ref{eq:efe}).
Also, writing a lagrangian density that includes gravitational 
terms and terms from quantum field theory leads to the 
temptation to combine some of these terms; however, one 
should bear in mind that we don't have such a unified 
theory yet.

Now, if one considers that a geometrical cosmological 
constant term is an integral part of the EFE 
then the old cosmological constant problem can 
be expressed  as (see for example 
\cite{review1,review6}): Why do all the contributions 
from vacuum energy density fantastically cancel with 
the geometrical $\Lambda$ term? 
This formulation of the problem seems to be less
often recalled in some of the recent literature.
To see this quantitatively, one can combine equations 
(\ref{eq:efe}) and (\ref{eq:vacT2}) to write
\begin{equation}
\label{eq:effective}
\frac{\Lambda_{effective}}{8\pi G}=\frac{\Lambda}{8 \pi G}+ \rho_{vac}
\end{equation}
where from effective quantum field theory 
\begin{equation}
\rho_{vac}=\frac{1}{2}\sum\hbar\omega.
\end{equation}
Now, if one considers only quantum field fluctuations cut off
at particle energies of 100 GeV (this means, we consider only
the well known physics of the standard model), one can 
write (with $\hbar=c=1$) $\rho_{vac} \sim (100 GeV)^4$.
However, from astrophysical observations $({\Lambda_{effective}}/{8\pi G})$
is found to be comparable to the critical density, i.e. 
$ \sim 10^{-48}$ $GeV^4$ which means that
the two terms in the RHS of (\ref{eq:effective}) must  
cancel to 56 decimal places.

On the other hand, from the identification of the 
cosmological constant with the vacuum energy, the old  $\Lambda$ 
 problem becomes: Why is the vacuum energy measured from 
astrophysics ($\sim 10^{-48}$ $GeV^4$) so small compared 
to the value of the vacuum energy estimated from quantum field 
theory ($\sim (100 GeV)^4$)? Of course, the situation is made 
even worse by taking the GUT or Planck scales.

This full identification may bring some limitations of 
its own to the old cosmological constant problem and 
is questioned later in this work.
\section{How much fine-tuning does the cosmic coincidence represent? (the new "problem")}
\label{sec:coincidence}
The cosmic coincidence problem is usually stated as why 
is the cosmic acceleration recent in the cosmic history of the 
universe (why ``now''?) 
It is also discussed in terms of a fine tuning problem: i.e. 
why is the matter energy density of the same order of 
magnitude as the dark energy density at the present epoch?
Related to this question is the fact that if the dark energy 
density was much larger then what is measured it would 
have dominated over the matter energy density much earlier and 
prevented structures from forming in the universe.
It is important to discuss the cosmic  coincidence because 
it is an argument that is used to motivate the search for
 dynamical components in order to be 
able to explain the coincidence.
We will try to quantify some of these statements here.

The matter energy density is related to the redshift by 
\begin{equation}
\label{eq:rho_z}
\rho_m(z)=(1+z)^3 \rho_m^{today}(z=0),
\end{equation}
and in a $\Lambda$CDM universe 
\begin{equation}
\label{age}
t(a)=\int^{a}_0\frac{da'}{a'H(a')}=\int^{a}_{0}\frac{da'}{a'H_0\sqrt{\Omega_{\Lambda}+\Omega_{m}a'^{-3}}}
\end{equation}
where $a=\frac{1}{1+z}$ is normalized to 1 today.

Now, from the concordance model $\Omega_m \approx 0.3$ and
 $\Omega_{\Lambda} \approx 0.7$ today and it follows that
\begin{equation}
\label{eq:rho_z_Today}
\rho_{\Lambda}=\frac{\Omega_{\Lambda}}{\Omega_m}\rho_m^{today}\approx
2.33 \rho_m^{today}.
\end{equation}
At the transition from deceleration to acceleration the dark energy
density is half the matter energy density and 
\begin{equation}
1+z_{trans}=\left(\frac{2\Omega_{\Lambda}}{\Omega_m}\right)^{1/3}\approx 1.67
\end{equation}
(see for a first measurement of this \cite{supernovae}:
$1+z_{trans}=1.46\pm0.13$). 

It is a clarifying exercise to evaluate the age of the universe 
$t$ at this transition. Using (\ref{age}) gives
a transition age of $\approx 7.14$ billion years (we use $H_0=72km/s/Mpc$ \cite{cmb}),
showing that the transition is not that recent in the time history 
of the universe. Moreover, $\rho_m$ and $\rho_{\Lambda}$ have been
within the same order of magnitude starting roughly from $1+z \approx 2.85$
(with $\rho_m \approx 9.94 \rho_{\Lambda}$) and that corresponds to an age of the
universe of $\approx 3.38$ billion years.
i.e. $\rho_m$ and $\rho_{\Lambda}$ have been within the same 
order of magnitude for at least the last 10.32 billions years
of the time history of the universe.

In order to evaluate how much fine-tuning the cosmic coincidence represents, 
one could evaluate some informative fractions using the transition time 
or the period during which dark energy has been of the same order 
of magnitude as the matter energy density, compared to the whole history
of the universe. These fractions provide the percent chances 
for an observer randomly put in the time history of the universe to 
have the matter energy density and the dark energy density being
of the same order of magnitude. 
We will compare energy densities using a logarithmic scale
of time, so the results will be similar to those using
a logarithmic redshift scale.

On a logarithmic scale, things can appear very
significant on plots; however, the fractions are actually of 
 a few percent:
\begin{equation}
\label{eq:probability1}
  \frac{\ln{t_{today}}-\ln{t_{transition}}}{\ln{t_{today}}-
  \ln{t_i}}\approx 2\%,
\end{equation}
\begin{equation}
\label{eq:probability2}
\frac{\ln{t_{today}}-\ln{t_{same-order}}}{\ln{t_{today}}-
  \ln{t_i}}\approx 3\%
\end{equation}
and
\begin{equation}
\label{eq:probability3}
\frac{\ln{t_{today}}-\ln{t_{same-order}}}{\ln{t_{today}}-
  \ln{t_{Planck}}}\approx 1\%
\end{equation}
We used here $t_i=1\,sec$ as we considered observational 
constraints from the Big Bang nucleosynthesis and CMB results 
to trace back in time the universe up to 1 second
 at the $e^{+}e^{-}$ annihilation. We also use 
 the Planck time $t_{Planck}=10^{-44}\;sec$ and find
that the fraction is $\approx 1\%$. 
(Of course on a linear scale of time the coincidence is insignificant: 
\begin{equation}
\frac{t_{today}-t_{transition}}{t_{today}-t_i}\approx 48 \%
\end{equation}
and 
\begin{equation}
\frac{t_{today}-t_{same-order}}{t_{today}-t_i}\approx 75 \%.)
\end{equation}

In summary, even on a logarithmic scale 
 the fractions remain of the order of 
a few percent.  
This is significant but not totally unusual in physics and astrophysics. 

Consequently, while it remains perhaps a motivated problem to seek 
models that could naturally explain some of these small fractions, 
this fine tuning should not constitute a barrier to reject 
models that address successfully the other cosmological 
constant problems.
\section{Observational constraints on the equation of state parameters and the cosmological constant/dark energy questions} 
An important approach to the cosmological constant questions is to parameterize the dark energy using its equation of state, and then to use different cosmological data sets in order to determine these parameters. One can thus infer answers to the cosmological constant/dark energy questions from the values found for these parameters. As we will delineate further, some results will be more conclusive than others. It is essential to combine multiple complementary cosmological probes and techniques in order to break degeneracies between the cosmological parameters including the dark energy parameter space. A powerful combination for constraining the dark energy parameters is the Cosmic Microwave Background Radiation (CMB) plus distance measurements from the type Ia Supernovae (SNIe) and Weak Gravitational Lensing (WL) also called cosmic shear, e.g. see 
\cite{2003ARA&A..41..645R,2001PhR...340..291B,1999ARA&A..37..127M}. These probes provide orthogonal constraints that, when combined, reduce significantly the uncertainties on the individual parameters. Also, constraints from clusters of galaxies and baryons oscillations were shown to be good probes of the dark energy equation of state, see e.g. \cite{mohr,wang2004}. Interestingly, WL allows one to do tomography by separating the source-galaxies into redshift bins in order to obtain further improvements on the parameter constraints, see e.g. \cite{Hu01,Takada,Ishak}. In this analysis, we consider future constraints from the CMB+SNIe+WL combination plus WL tomography with various numbers of redshift bins.  
\subsection{methodology}
\label{sec:methodology}
\begin{table}
{
\caption{Cosmological model. We use fiducial values for the parameters 
from recent results from the WMAP (3-years data results) + Large scale Structure + Supernova data 
as listed in table \ref{tab:surveys} with the respective references (we use for the equation of state the cosmological constant parameter values). 
We assume a spatially flat universe with $\Omega_{m}+\Omega_{\Lambda}=1$. 
This fixes $\Omega_m$ and $H_0$ as functions of the basic parameters. 
We do not include massive neutrinos, or primordial isocurvature perturbations.}
\label{tab:fiducial}
\begin{tabular}{|l|l|l|l|l|l|l|}
\hline
Symbol                & Description                                                  & Fiducial value&Probe-1&Probe-2&Probe-3\\\hline
$\Omega_{m}h^2$       & physical matter density                                      &  0.1259       & WL& CMB& SNIe\\
$\Omega_\Lambda$      & fraction of the critical density in a dark energy component  &  0.745        & WL& CMB& SNIe\\
$w_0$,$w_1$ (or $w_a$)& equation of state parameters                                 &  -1, 0        & WL& CMB& SNIe\\
$\sigma_8^{\rm lin}$  & amplitude of linear fluctuations                             &  0.712        & WL& CMB&   \\
$n_s(k_0=0.05h/{\rm Mpc})$&spectral index of the primordial scalar power spectrum at $k_0$&  0.946   & WL& CMB&   \\
$\alpha_s$            &  running of the primordial scalar power spectrum             & -0.06         & WL& CMB&   \\
$z_p$                 & the characteristic redshift of source galaxies for lensing   & 0.76, 1.12    & WL&    &   \\
$\cals$               & absolute calibration parameter for WL power spectrum \cite{2003MNRAS.343..459H,Ishak2004}& 0.0 & WL&&\\
$\calr$               & relative calibration parameter for WL power spectrum \cite{2003MNRAS.343..459H,Ishak2004}& 0.0 & WL&&\\
$\Omega_{b}h^2$       & physical baryon density                                  & 0.0448       & & CMB &\\
$\tau$                & optical depth to reionization                            & 0.08         & & CMB &\\
$T/S$                 & scalar-tensor fluctuation ratio                          & 0.2          & & CMB &\\
\hline
\end{tabular}
}
\end{table}
In this analysis, we consider a cosmological model with 13 parameters that take the 
fiducial values in table \ref{tab:fiducial} based on recent 3-years results from the 
Wilkinson Microwave Anisotropy Probe (WMAP) \cite{Bennet2003,Spergel2003, Spergel2006} 
combined with large scale structure and supernova data as given in table \ref{tab:surveys} 
below along with the respective references (except for the dark energy equation of state, 
where we use the cosmological constant values). We consider in this analysis, the two standard 
parameterizations for the dark energy equation of state $w = P / \rho$ given by  
\begin{itemize}
\item{$w(z) = w_0 + w_1 z$ if $z<1$ and $w(z)= w_0 + w_1$ otherwise, e.g.  
\cite{supernovae,Upadhye}, and}
\item{$w(a)= w_0 + w_a (1-a)$, e.g. \cite{Chevallier,Linder}.} 
\end{itemize}
The dark energy density as a function of redshift is thus given by  
\begin{equation}
{\mathcal{Q}}(z) \equiv \rho_{de}(z) / \rho_{de}(0)=\exp{3\int_{0}^{z}\frac{1+w(z'))dz'}{1+z'}}.
\label{eq:de_function}
\end{equation} 
Despite its simplicity, statistical inference theory using Fisher matrices was 
proven to be a very efficient tool to calculate constraints that will be obtained
from future planned or proposed experiments. As an example of its efficiency, 
one could look at the good predictions on the parameter uncertainties forecasted, 
for instance, by \cite{Wang1999,Hu01} for WMAP and compare them to the real WMAP results, 
obtained years later \cite{Spergel2003}. We describe here only the basic ideas of the 
methods that we use in the current paper but provide references \cite{fisher} where 
the full detail can be found.

We use the standard approach to calculate the statistical error on a given parameter 
$p^{\alpha}$ by using the combination:  
\begin{equation}
\sigma^2(p^{\alpha})\approx [({\bf F}_{CMB}+{\bf F}_{WL}+{\bf F}_{SNe}+\Pi)^{-1}]^{\alpha \alpha},
\label{eq:sigma}
\end{equation}
where ${\bf F}_{CMB}$, ${\bf F}_{WL}$ and ${\bf F}_{SNe}$ are the Fisher matrices 
from CMB, weak lensing, and supernovae respectively, and $\Pi$ is the prior matrix. 
As discussed in some detail in \cite{Hu01,Ishak}, we calculate ${\bf F}_{WL}$ using 
\begin{equation}
F_{\alpha \beta} = {\sum_{\ell=\ell_{\rm min}}^{\ell_{\rm
        max}}{\frac{1}{(\Delta P_\kappa
        )^2} {\frac{\partial P_\kappa}{\partial p^{\alpha}}}}
        {\frac{\partial P_\kappa}{\partial p^{\beta}}}};
\label{eq:fisher1}
\end{equation}
where the lensing convergence power spectrum is given by \cite{1992ApJ...388..272K,1997ApJ...484..560J,1998ApJ...498...26K}:
\begin{equation}
P^{\kappa}_l =
\frac{9}{4} H_0^4\Omega_m^2\int^{\chi_H}_{0}
\frac{g^2(\chi)}{a^2(\chi)}P_{3D}
\left(\frac{l}{\sin_{K}(\chi)},\chi\right) d\chi.
\end{equation}
\begin{table*}
{
\begin{center}
\caption{Data sets used in \cite{Spergel2006} to derive the constraints on equation of state parameter w using CMB, large scale structure and supernova data (table 9 in \cite{Spergel2006}). See also the NASA's data center for Cosmic Microwave Background (CMB) research, LAMBDA at http://lambda.gsfc.nasa.gov/.
}
\label{tab:surveys}
\begin{tabular}{|l|l|}
\hline
data set/experiment                      &                   references \\\hline
WMAP 3-years results                     & (Spergel, et al., 2006) \cite{Spergel2006}\\  
2dF Galaxy Redshift Survey               & (Cole, et al., 2005) \cite{Cole2005} \\
BOOMERanG + ACBAR                        & (Montroy, et al., 2005, Kuo, et.al., 2004) \cite{Montroy2005,Kuo2004}\\
CBI + VSA                                & (Readhead, et al., 2004, Dickinson, et.al., 2004) \cite{Readhead2004,Dickinson2004}\\
Sloan Digital Sky Survey                 & (Tegmark, et al., 2004; Eisenstein, et.al., 2005) \cite{Tegmark2004,Eisenstein2005}\\
Supernova "Gold Sample"                  & (Riess, et.al., 2004) \cite{Riess2004}\\
Supernova Legacy Survey (SNLS)           & (Astier, et al., 2006) \cite{Astier2006}\\

\hline
\end{tabular}
 \end{center}
}
\end{table*}

where $P_{3D}$ is the $3D$ nonlinear power spectrum of the matter
density fluctuation, $\delta$;  $a(\chi)$ is the scale factor; and
$\sin_{K}\chi=K^{-1/2}\sin(K^{1/2}\chi)$ is the comoving angular
diameter distance to $\chi$ (for the spatially flat universe used in
this analysis, this reduces to $\chi$). The weighting function
$g(\chi)$ is the source-averaged distance ratio given by
\begin{equation}
\label{eq:weighting}
g(\chi) = \int_\chi^{\chi_H} n(\chi') {\sin_K(\chi'-\chi)\over
\sin_K(\chi')} d\chi',
\end{equation}
where $n(\chi(z))$ is the source redshift distribution normalized by
$\int dz\; n(z)=1$. The uncertainty in the observed lensing spectrum is
given by: \cite{1992ApJ...388..272K,1998ApJ...498...26K}
\begin{equation}
\Delta P_{\kappa}(\ell)=
\sqrt{\frac{2}{(2\ell +1)f_{sky}}}\left (
P_{\kappa}(\ell) + {\left< \gamma_{int}{}^2 \right>\over {\bar n}} \right ) \,,
\label{delta_kappa}
\end{equation}
\noindent where $f_{sky} = \Theta^2 \pi/129600 $ is the fraction of the
sky covered by the gravitational lensing survey of dimension $\Theta$ in 
degrees, and $\left<\gamma_{int}^2\right>^{1/2}$ is the intrinsic ellipticity 
of galaxies. For this analysis, we consider a lensing survey with $10\%$ 
sky coverage, a median redshift of 1, an average galaxy number 
density of $\bar n = 30\; {\rm gal/arcmin}^2$, and intrinsic ellipticities 
$\left<\gamma_{int}^2\right>^{1/2}=0.4$. We calculate the constraints 
on the parameters using five tomographic bins for this survey. 
We also consider a deeper survey (space-based like) with $f_{sky}=0.10$, 
a median redshift of roughly 1.5, $\bar n = 100\;{\rm  gal/arcmin}^2$, 
and $\left<\gamma_{int}^2\right>^{1/2} \approx 0.25$. We calculate 
constraints from 10 tomographic bins using this survey. For both 
surveys, we use $\ell_{\rm max}=3000$ to keep the assumption of a 
Gaussian shear field valid. 

Weak lensing has been recognized as a very powerful probe of dark energy 
parameters, however, several systematic effects have been identified 
so far, see \cite{2003ARA&A..41..645R} and references therein for an overview. 
In this analysis, we included the effect of the shear calibration bias 
\cite{Erben,Bacon,2003MNRAS.343..459H,2002AJ....123..583B,2000ApJ...537..555K,2003astro.ph..5089V} 
on our results by marginalizing over its parameters. Because of this bias, the
shear is systematically over or under-estimated by a multiplicative
factor, and results in an overall rescaling of the shear power
spectrum. Following the parameterization discussed in \cite{Ishak2004}, 
we used the absolute power calibration parameter $\cals$
and the relative calibration parameter $\calr$ between two redshift
bins. Another systematic effect that we considered in the analysis is the 
incomplete knowledge of the source redshift distribution \cite{2000Natur.405..143W,
IshakHirata}. A remedy to this poor knowledge of the redshift distribution using 
spectroscopic redshift has been explored recently in \cite{IshakHirata}. In the 
present analysis we marginalize over the redshift bias by including the 
characteristic redshift of the distribution as a systematic parameter $z_s$ and 
we assume a reasonable prior of 0.05 on this parameter. 

In order to calculate the supernova Fisher matrix, ${\bf F}_{SNe}$, we 
use (see, e.g. \cite{snfisher,HutererTurner})
\begin{equation}
F_{\alpha \beta} = {\sum_{i=1}^{N}}
        {\frac{1}{\sigma_m(D_{L,i})^2}
        {\frac{\partial D_{L,i} }{\partial p^{\alpha}}}}
        {\frac{\partial D_{L,i}}{\partial p^{\beta}}}.
\label{eq:snfisher}
\end{equation}
where $D_L\equiv H_0 d_L/c $ is the dimensionless luminosity
distance to a supernova, given in a spatially flat model by
\begin{equation}
D_L(z) = (1+z) \; \int_0^z \frac{1}{\sqrt{(1-\Omega_\Lambda)(1+z')^3 +
    \Omega_\Lambda\mathcal{Q}(z')}} dz',
\label{eqn:DL_vs_z}
\end{equation}
where  $\mathcal{Q}(z)$ is as defined in Eq.(\ref{eq:de_function}). 
We recall that the SN Ia apparent magnitude as a function of redshift is 
given by $m(z) = 5 \log_{10} \left( D_L(z) \right) + {\mathcal{M}}$, where
${\mathcal{M}}$ depends as the absolute magnitude of type Ia supernovae
as well as on the Hubble parameter $H_0$. As usual, we treat ${\mathcal{M}}$ 
as a nuisance parameter. We use two sets of 2000 SNe Ia uniformly distributed 
with $z_{max}=0.8$ and $z_{max}=1.5$.  It is important to briefly note here
that there are systematic uncertainties associated with supernova
searches: these include  luminosity evolution, gravitational lensing
and dust; see, e.g. \cite{SNAP2003b} and references therein. In order
to partly include the effect of these systematics and the effect of
the supernova peculiar velocity uncertainty \cite{Tonry}, we follow
\cite{Kim2003,SNAP2003b} and use the following quadrature for the effective
magnitude uncertainty 
\begin{equation}
\sigma_m^{eff}=\sqrt{\sigma_m^2+\left(\frac{5\sigma_v}{ln(10)cz}\right)^2+N_{\{per\;bin\}}\delta_m^2}
\label{eq:snquadrature}
\end{equation}
where, $\sigma_v=500km/sec$ is the peculiar velocity, and
$\delta_m$ is a floor uncertainty in each bin \cite{Kim2003,SNAP2003b}. 
The quadrature relation (\ref{eq:snquadrature}) assures that there is an 
uncertainty floor set by the systematic limit $\delta_m$ so that the overall 
uncertainty per bin cannot be reduced to arbitrarily low values by adding 
more supernovae.

Finally, for the CMB and lensing tomography, we use the generalized form of the Fisher matrix above (e.g. see \cite{Hu01}) as  
\begin{equation}
 F_{\alpha\beta} = \sum_{\ell_{\rm min}}^{\ell_{\rm max}}
                (\ell + 1/2) f_{\rm sky}
 {\rm Tr}\left( {\bf C}_\ell^{-1} {\partial{\bf C}_\ell\over\partial p^\alpha}
                       {\bf C}_\ell^{-1} {\partial{\bf C}_\ell\over\partial p^\beta} \right),
\label{eq:fisher2}
\end{equation}
where ${\bf C}_\ell$ is the covariance matrix of the multipole moments
of the observables ${C}^{X Y}_\ell = C_\ell^{X Y} + N_\ell^{X Y}$ with 
$N_\ell^{X Y}$ being the power spectrum of the noise in the measurement. Here $X Y$ 
takes the values $\kappa \kappa$ for lensing tomography spectra and cross-spectra, and $T T$, $T E$, 
and $E E$ for CMB spectra. We consider constraints from 1-year data from the Planck satellite.
Our findigs are discussed in the next section.
\subsection{Summary of results and implications for the cosmological constant/dark energy questions}
\begin{table*}
{
  \begin{center}
  \caption{Dark energy parameter constraints from various combinations of CMB, Weak 
Gravitational Lensing and Supernova future surveys. The constraints are ($1\sigma$ uncertainties)
on the two dark energy parameterizations given in section \ref{sec:methodology}. The uncertainties 
are calculated using combinations of 1-year data from Planck, 2000 uniformly distributed supernovae 
with $z_{max}=0.8, 1.5$, a lensing survey with 5-bins tomography, and a very 
deep lensing survey with 10-bins tomography. The results are presented for the 
dark energy parameters $\{w_0, w_1\}$ and $\{w_0, w_a\}$. The systematic effects discussed 
in section \ref{sec:methodology} are included in the calculations. 
}
\label{tab:results}
  \begin{tabular}{|l|c|c|c|c|}
\hline
Simulated Experiment/Survey    &\multicolumn{2}{c|}{Parameterization-1}&\multicolumn{2}{c|}{parameterization-2}\\
\cline{2-5}
                                                              & \,\,\,$\sigma(w_0)$\,\,\, &$\sigma(w_1)$ & \,\,\,$\sigma(w_0)$\,\,\, & $\sigma(w_a)$\\\hline

1-year data from PLANCK + 2000 SN with $z_{max}=0.8$&  0.11        &  0.26        &  0.13        &  0.47         \\ 
                                                &             &              &              &               \\\hline
1-year data from PLANCK + 2000 SN with $z_{max}=0.8$&  0.039      &  0.092       &  0.033       &  0.12         \\
+ WL survey with 5-bins tomography, $z_{med}=1.0$ and $f_{sky}=0.10$ &              &              &             &               \\\hline
1-year data from PLANCK + 2000 SN with $z_{max}=1.5$& 0.08        &  0.20        &   0.10       & 0.35         \\
                                                  &             &              &              &               \\\hline
1-year data from PLANCK + 2000 SN with $z_{max}=1.5$&   0.022      &   0.043     &   0.024      &    0.057      \\
+ WL deep survey with 10-bins tomography, $z_{med}=1.5$ and $f_{sky}=0.10$&             &              &             &               \\
\hline
  \end{tabular}
 \end{center}
}
\end{table*}
Our results, summarized in table \ref{tab:results}, show that when WL 
with multiple-bins tomography is added to the CMB+SN combination, the uncertainties 
on the equation of state parameter $w_0$ reduce by roughly a factor of 3 and the 
uncertainties on $w_1$ and $w_a$ reduce by roughly factors of 3 or 4 at least.
Also, we note that the uncertainties on the set $\{w_0,w_1\}$ are different from those 
of the set $\{w_0,w_a\}$ showing that the uncertainties are parameterization dependent. 
The deeper lensing survey, which allows one to implement 10-bins tomography, 
provides an additional factor of 2 improvement on $w_1$ and $w_a$ compared with 
the 5-bins tomography results. We find that in order to bring the uncertainties 
on both parameters to the order of several percent, very ambitious surveys are required. 
Current observations are only able to constrain, with some significance, models with a 
constant equation of state, i.e. one single parameter with no redshift dependence. 
For example, the WMAP team recently combined constraints from currently available 
CMB, large scale structure, and supernova data sets (see table \ref{tab:surveys}) and 
obtained $w=-0.926^{+0.051}_{-0.057}$ (table 9 in \cite{Spergel2006}).

Our table \ref{tab:results} shows the 1-sigma uncertainties on the equations of state 
parameters, and in order to consider the 2 and 3-sigma constraints, one has to multiply the 
values found by factors of two and three respectively. Therefore, one can see that 
even when ambitious surveys are considered the remaining uncertainty is still too 
large to constrain significantly multiple-parameters dark energy models. However, 
It will be possible to exclude some proposed models with significant 
deviations from the cosmological constant parameters, namely $w_0=-1$, and $w_1=0$. 
These include for example trackers models \cite{zlatev} and some SUGRA inspired models 
\cite{brax} with for example $w_0=-0.8$ and  $w_1=0.3$. 
The most decisive answer will be if the data can show conclusively that dark energy is not 
a cosmological constant. A very suggestive but less decisive answer will be to show that the 
dark energy parameters are those of a cosmological constant to a very high level of precision 
(a few percent, perhaps). But of course, the degeneracy in this case will remain and other 
tests, beyond the equation of state approach, will be needed. In this case, other tests are 
also necessary because the problems of the cosmological constant are just re-affirmed. 
Finally, an important question that need to be addressed in all cases is whether the equation 
of state obtained is a true or forced equation of state as we explore further below. We discuss 
in the next sections some possible directions of such tests and illustrate one promising test using cosmological probes.
\section{Reconsidering and testing some of the assumptions about the old cosmological constant and dark energy problems} \label{sec:solution}
In this section we propose some examples on how one could question 
some of the assumptions made about the cosmological constant/dark energy  
problems and, in some cases, how to put this questioning to the test. 
\subsection{Cosmological tests beyond the equation of state approach: 
The expansion history versus the growth rate of large scale structure}

The approach of the equation of state is certainly an important one. However, 
an important question that remains after some dark energy parameters are obtained 
from analyzing observational data is as follows. Is this an effective equation 
of state of some dark energy component in the Einstein's equations, or is this 
just a forced equation of state obtained from fitting dark energy models on the 
top of some modified gravity at cosmological scales? New tests are necessary in 
order to address this question. 

Indeed, of great importance are innovative ways of using
current and future  astrophysical observations that could 
distinguish between acceleration models beyond the equation of state. 
In other words, for the same degenerate effective equation of state, these 
novel tests could distinguish between dynamical dark energy 
models, a geometrical cosmological constant, and acceleration due to some 
modification to the gravity sector, and thus will allow one to test some of 
the important basic assumptions. Some cosmological probes such as weak gravitational lensing (for reviews, 
see \cite{2003ARA&A..41..645R,2003astro.ph..5089V,2001PhR...340..291B,
1999ARA&A..37..127M} and references therein) and clusters of galaxies 
(see for example \cite{mohr,wang2004} and references therein) are very rich 
tools and very promising for identifying this type of test because they 
provide more than one way to constrain dark energy or cosmic acceleration.
Both probes can capture the effect of dark energy on the expansion 
history and also its effect on the growth rate of large-scale structure
(the rate at which clusters and super clusters of galaxies form over 
the history of the universe). Interestingly, this can be used to identify 
consistency checks to test the dark energy beyond the equation of state 
degeneracy. In particular, this could allow one to test dark energy models
based on new particles and fields versus cosmic acceleration due to some 
modification in the curvature sector of the EFE as suggested in some recent 
studies, see e.g. \cite{carroll2,capozziello1,dgp}, recent review \cite{review8} 
and references therein. Most importantly, these kinds of consistency checks 
could be used to test some of the assumptions discussed in this paper, 
namely on the origin of the intrinsic constant curvature of spacetime. 
Another test beyond the equation of state and based on the dark energy potential 
was discussed in \cite{Jiminez}. 

In this paper, we explore an example to demonstrate that tests that go 
beyond the equation of state degeneracy are possible. An important point 
for this test is that cosmic acceleration affects cosmology in two ways: 
{\it 1)} It affects the expansion history of the universe by speeding it up,
{\it 2)} It affects the growth rate of large scale structure in the universe 
 by suppressing it. The idea explored is that, for dark energy models, 
these two effects must be consistent one with another because their 
respective functions are mathematically related by General Relativity equations. 
The presence of significant inconsistencies between the expansion 
Hubble function and the growth rate function could be the signature of 
some modified gravity at cosmological scales as we will demonstrate. 

In order to illustrate how the test works, we will need to use a viable 
modified gravity model. We choose a model proposed by Dvali, Gabadadze 
and Porrati (DGP) \cite{dgp} where the cosmic acceleration is due to 
the effect of an extra large dimension modifying gravity at cosmological scales. 
This DGP model is motivated by higher dimensional physics and is not ruled 
out by current astrophysical observations \cite{DeffayetLandauRaux,lue,LueReport}.
We have no particular interest in the phenomenology and precise testing of 
the viability of this model. We are only interested to use it as an example 
in order to illustrate the test considered. We refer the interested reader 
to some studies dedicated to the DGP model phenomenology 
\cite{Deffayet3,Sawicki,LueReport,Fairbairn}.

We provide here a very brief description of this model but again refer the reader 
to \cite{dgp,DeffayetDvaliGabadadze2002} for a full description. The action for
this five-dimensional theory is \cite{dgp, DeffayetDvaliGabadadze2002}
\begin{equation}
S_{(5)}=\frac{1}{2}M_{(5)}^3 \int d^4x \, dy \sqrt{-g_{(5)}}
R_{(5)}+\frac{1}{2}M_{(4)}^2 \int d^4x \sqrt{-g_{(4)}} R_{(4)}
+ S_{matter}, 
\end{equation}
where the subscripts $4$ and $5$ denote quantities on
the brane and in the bulk, respectively; $M_{(5)}$ is the five
dimensional reduced Planck mass; $M_{(4)}=2.4 \times 10^{18}
\textrm{GeV}$ is the four dimensional effective reduced Planck mass;
$R$ and $g$ are the Ricci scalar and the determinant of the metric,
respectively.  The first and second terms on the right hand side
describe the bulk and the brane, respectively, while $S_{matter}$ is
the action for matter confined to the brane. 
The two different prefactors $M_{(5)}^3/2$ and $M_{(4)}^2/2$ in front
of the bulk and brane actions give rise to a characteristic length
scale \cite{Deffayet2001}, $r_c = {M_{(4)}^2}/{2 M_{(5)}^3}$. 
 If $M_{(5)}$ is much less than $M_{(4)}$, then the brane terms in the
action above will dominate over the bulk terms on
scales much smaller than $r_c$, and gravity will appear four
dimensional.  On scales larger than $r_c$, the full five dimensional 
physics will be recovered, and the gravitational force law will revert 
to its five dimensional $1/r^3$ form.  This is usually discussed in terms 
of gravity leakage into an extra dimension. Ref. \cite{DeffayetLandauRaux} 
shows that tuning $M_{(5)}$  to about $10-100 \textrm{MeV}$, 
implying $r_c \sim H_0^{-1}$, is consistent with cosmological data. 
They have also been discussed in \cite{lue}. 
Low redshift cosmology in DGP brane worlds was studied in
\cite{Deffayet2001,DeffayetDvaliGabadadze2002}. 
Following \cite{DeffayetDvaliGabadadze2002}, one could define the effective 
energy density $\rho_{r_c} \equiv \frac{3}{(32 \pi G r_c^2)}$ so that  
Friedmann's first equation becomes 
\begin{equation}
H_{_{DGP}}^2 + \frac{k}{a^2} = \frac{8 \pi
  G}{3} \left(\sqrt{\rho + \rho_{r_c}} + \sqrt{\rho_{r_c}}
\right)^2.
\end{equation}
where $\Omega_{r_c} \equiv \frac{1}{4} r_c^{-2} H_0^{-2}$. 
We focus here on a flat universe ($k=0$) containing only nonrelativistic
matter, such as baryons and cold dark matter, in which 
$\Omega_{r_c} = \left( \frac{1-\Omega_m}{2} \right)^2$.
In this model, the gravitational ``leakage'' into the fifth 
dimension, on large length scales, becomes a substitute for 
dark energy.  

Now, for the standard general relativistic (GR) cosmological model, FLRW, with zero spatial 
curvature (k=0), and a dark energy component (DE), the expansion history is expressed by the Hubble 
function and is given by 
\begin{equation}
H_{_{GR+DE}}(z) = H_0
\sqrt{\Omega_m (1+z)^3 + (1-\Omega_m) {\mathcal{Q}}(z)}.  
\label{eq:expansion_gr}
\end{equation}
And the growth rate of large scale structure G(a=1/(1+z)) is given by
integration of the differential equation, \cite{Ma1999,linder2003},
\begin{equation}
G_{_{GR+DE}}''+\left[\frac{7}{2}-\frac{3}{2}\frac{w(a)}{1+X(a)}\right]\frac{G'_{_{GR+DE}}}{a}+\frac{3}{2}\frac{1-w(a)}{1+X(a)}\frac{G_{_{GR+DE}}}{a^2}=0,  
\label{eq:growth_gr}
\end{equation}
where $'\equiv d/da$, $G=D/a$ is the normalized growth rate, $\mathcal{Q}(a)$ is as given by equation \ref{eq:de_function}, and 
\begin{equation}
X(a)=\frac{\Omega_m}{(1-\Omega_m)a^3 \mathcal{Q}(a)}. 
\end{equation}
\begin{figure}
\begin{center}
\begin{tabular}{cc}
{\includegraphics[width=2.5in,height=3.2in,angle=-90]{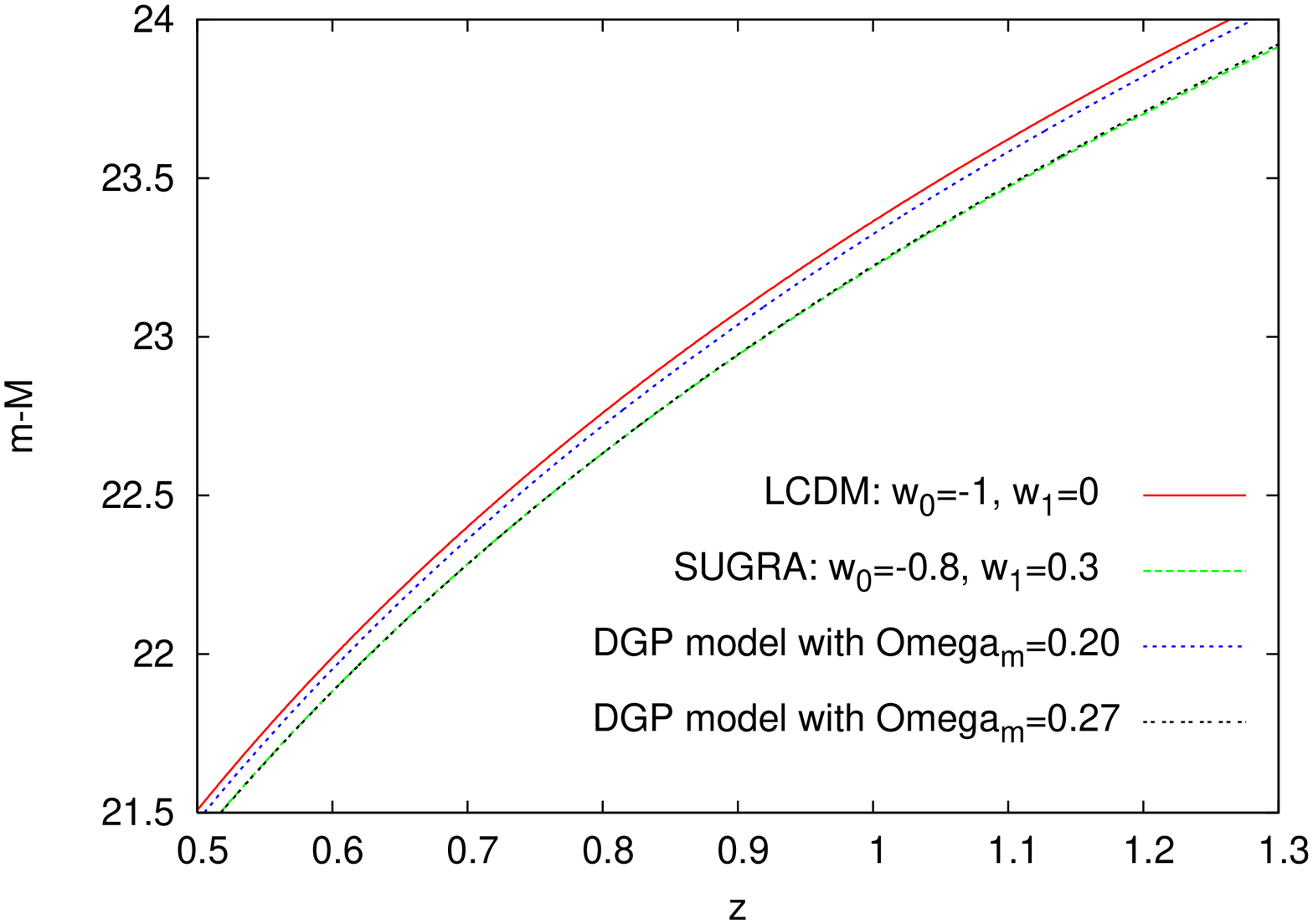}} &
{\includegraphics[width=2.5in,height=3.2in,angle=-90]{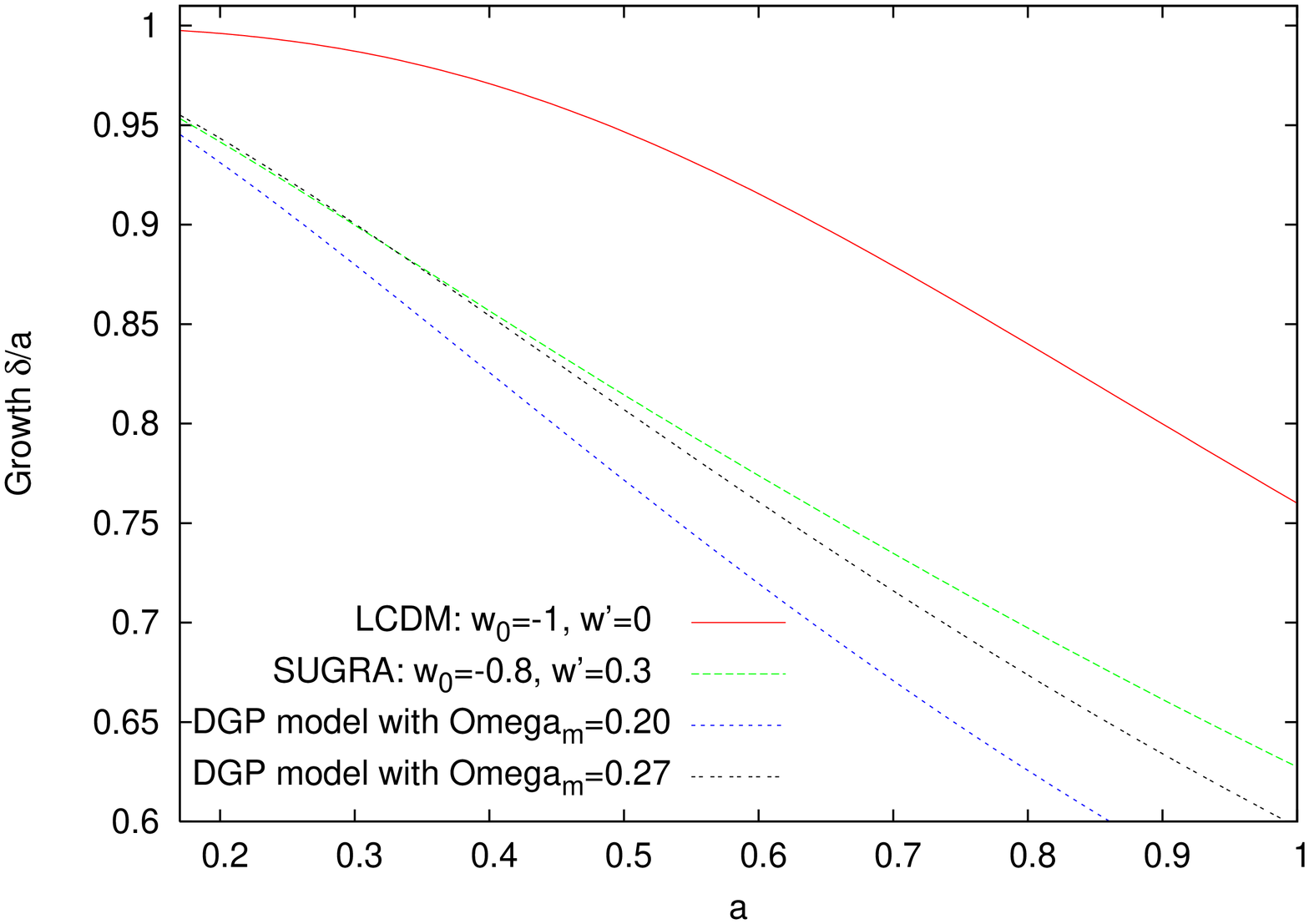}} \\
\end{tabular}
\caption{\label{fig:hubblesandgrowths} 
Supernova Hubble diagrams (Left) and Growth rate functions for dark energy and DGP models. 
 Note that the $\Lambda$CDM model (red solid line) and the $\Omega_m=0.20$ DGP 
model (blue dotted) have degenerate Hubble diagrams, but different 
growth rates.  The degeneracy of the Hubble diagrams is even stronger for SUGRA 
(green dashed) and $\Omega_m=0.27$  DGP (black double dotted) models. 
Interestingly, the growth rate in the $\Omega_m=0.27$ DGP model is suppressed 
with respect to that in the $\Lambda$CDM model, which has the same $\Omega_m$.} 
\end{center}
\end{figure}

On the other part, for the spatially flat DGP model, the expansion Hubble function is given by  
\begin{equation}
\label{eq:expansion_dgp}
{{H_{DGP}}}(z) = H_0\Big{(} \frac{1}{2}(1-\Omega_m)+ \sqrt{\frac{1}{4}(1-\Omega_m)^2 + \Omega_m (1+z)^3} \Big{)}. 
\end{equation}
and the suppression of the growth rate function is given by, \cite{lue,Koyama}
\begin{equation}
\ddot{\delta}_{_{DGP}}+2 H_{_{DGP}} \dot{\delta}_{_{DGP}}- 4 \pi G \rho \big{(}1+\frac{1}{3\beta}\big{)} \delta_{_{DGP}} =0
\label{eq:growth_dgp}
\end{equation}
where 
\begin{equation}
\beta=1-2r_c H_{_{DGP}} \Big{(}1+\frac{\dot{H}_{_{DGP}}}{3 H_{_{DGP}}^2}\Big{)}
\end{equation}
\indent
We illustrate in Fig.\ref{fig:hubblesandgrowths}, Hubble expansion functions and 
growth rate functions for dark energy and DGP models including the 
cosmological constant model, $\Lambda$CDM. The figure shows that compared 
to a $\Lambda$CDM model with the same matter density, a DGP model has a 
distinct suppression of the growth rate. Also, Fig.\ref{fig:hubblesandgrowths}b 
displays how the degenerate models of Fig.\ref{fig:hubblesandgrowths}a show distinct 
growth rate functions. 

The basic idea for the test explored is that equations (\ref{eq:expansion_gr})
and (\ref{eq:growth_gr}) must be mathematically consistent one with another 
via General Relativity. Similarly, equations (\ref{eq:expansion_dgp}) and 
(\ref{eq:growth_dgp}) must be consistent one with another via DGP theory. 
A consistency cross-check of these two functions constitute a test for 
the fundamental underlying theory.

In order to apply the test discussed below, we assume the availability 
of a sample of $N_{SNe}=2000$ type Ia supernova, evenly distributed in 
redshift between $z_{Min}=0$ and $z_{Max}=1.7$, with a magnitude uncertainty 
per supernova of $\sigma_m=0.2$.  As discussed in section \ref{sec:methodology}, 
we also include systematic effects using the quadrature (\ref{eq:snquadrature}). We also 
assume the availability of measurements of the growth rate in twenty bins 
evenly spaced in the scale factor $a$, between $a_{min}=0.25$ and $a_{max}=1$
and an uncertainty of $\sigma_{_G(a)}/G(a) = 0.02$. In order to break the 
usual degeneracy between $\Omega_m$ and the equation of state parameters,
we add to the SN Ia data and to the growth rate measurements a constraint from 
the CMB shift parameter. The CMB shift parameter for a spatially flat universe 
is given by  ${\mathcal{R}} = \Omega_m^{1/2} \int_0^{z_{_CMB}} \frac{dz}{{{H/H_0}}(z)}$, 
where $z_{CMB}=1089$ is the redshift of the surface of last scattering, given in
\cite{Spergel2003}. Ref. \cite{WangTegmark} reports that the value of
$\mathcal{R}$ from current data is ${\mathcal{R}}_{obs}=1.716 \pm 0.062$.  

In order to demonstrate the working of the consistency check we 
proceed as follows: 
We assume that the true cosmology is described by a DGP model and
simulate the expansion and growth data using a fiducial DGP model.  
Then we ask what contradictions arise when the data are instead analyzed 
based on the assumption of a dark energy model (as mentioned earlier, we are 
not particularly interested in the DGP cosmology here but we want to use it 
as an example to illustrate the procedure.) 

Because we will generate the data using the DGP model, the consistency relation 
from General Relativity between the expansion history and the growth rate of 
large scale structure will be broken. The dark energy equation 
of state $w_{exp}(z)$ which best fits measurements of the expansion 
 will not be consistent with the equation of state $w_{growth}(z)$ which 
best fits measurements of the growth. 

The methods and steps we use are as follows:
\begin{enumerate}
\item[i)]{We use a fiducial DGP model with $\Omega_m=0.27$ and 
simulate the data for the expansion and the growth rate. We generate 
supernova magnitudes, a growth rate function, and the CMB shift parameter}
\item[ii)]{We use a $\chi^2$ minimization method in order to
find the best fit dark energy model to the expansion. We obtain  
the best fit model to the supernova magnitudes and the CMB shift 
parameter with a first dark energy parameter space $\{\Omega_{de},w_0,w_1\}$. 
The $\chi^2$ minimization method was discussed in detail in \cite{Upadhye} and was 
shown to give similar results to those from Monte Carlo Markov Chain 
method although the $\chi^2$ minimization can have a better handle on 
degeneracies \cite{Upadhye})}
\item[iii)]{We use the $\chi^2$ minimization in order to
find the best fit dark energy model to measurements of the growth rate of large 
scale structure. We determine the best fit model to the growth rate function and the 
CMB shift parameter and obtain a second dark energy parameter space $\{\Omega_{de}',w_0',w_1'\}$} 
\item[iv)]{Next, we use a standard Fisher matrix approach to calculate 
the confidence regions (or  $\chi^2$ contours) around
the two best fit dark energy models (this standard procedure 
is discussed in detail in references \cite{fisher})}
\item[v)]{Then, we compare the allowed regions 
in the $\{\Omega_{de},w_0,w_1\}$ parameter space between the two 
data combinations in order to look for inconsistencies} 
\end{enumerate}

\begin{figure}
\begin{center}
\begin{tabular}{cc}
{\includegraphics[width=2.5in,height=3.2in,angle=-90]{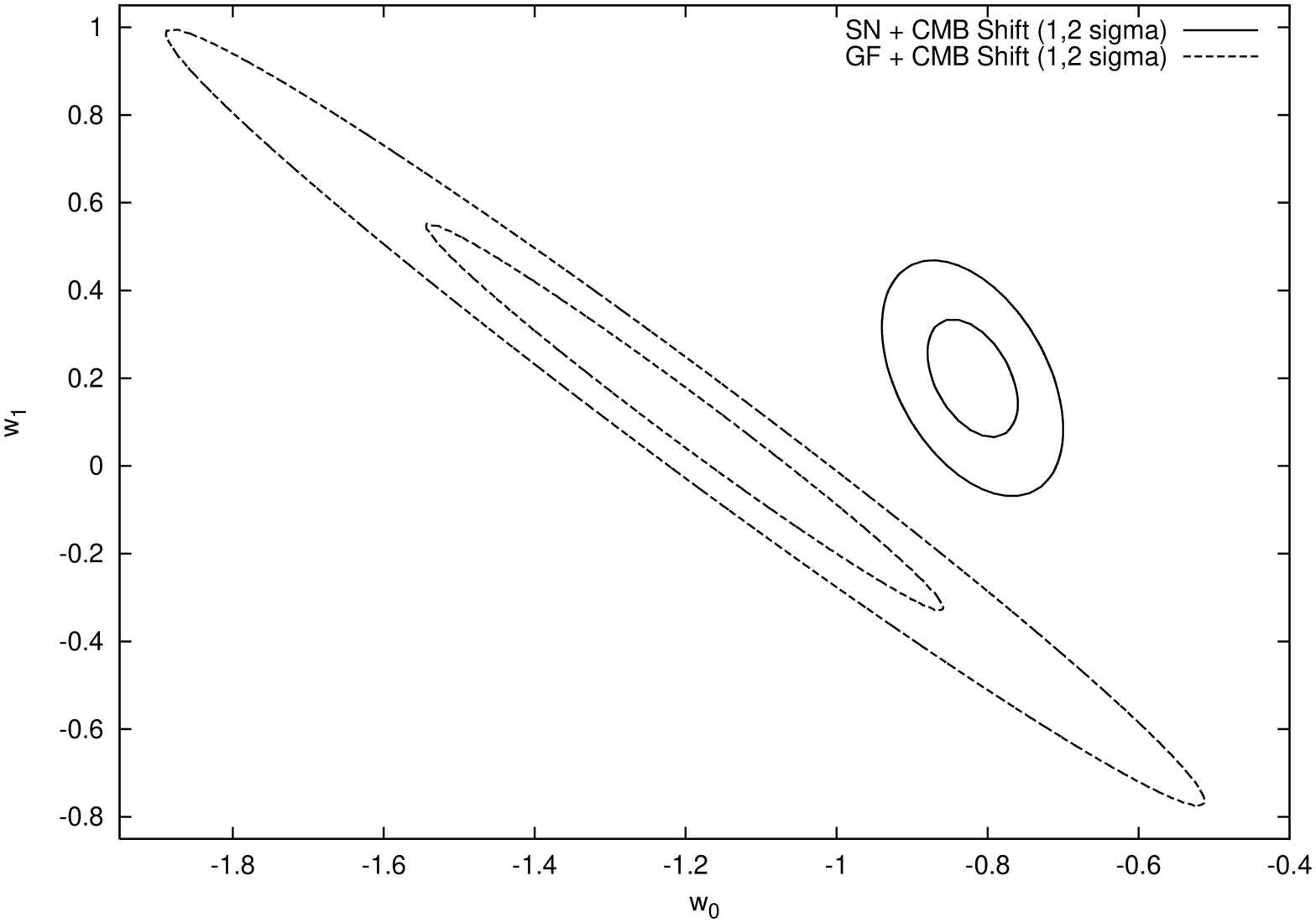}} &
{\includegraphics[width=2.3in,height=3.0in,angle=-90]{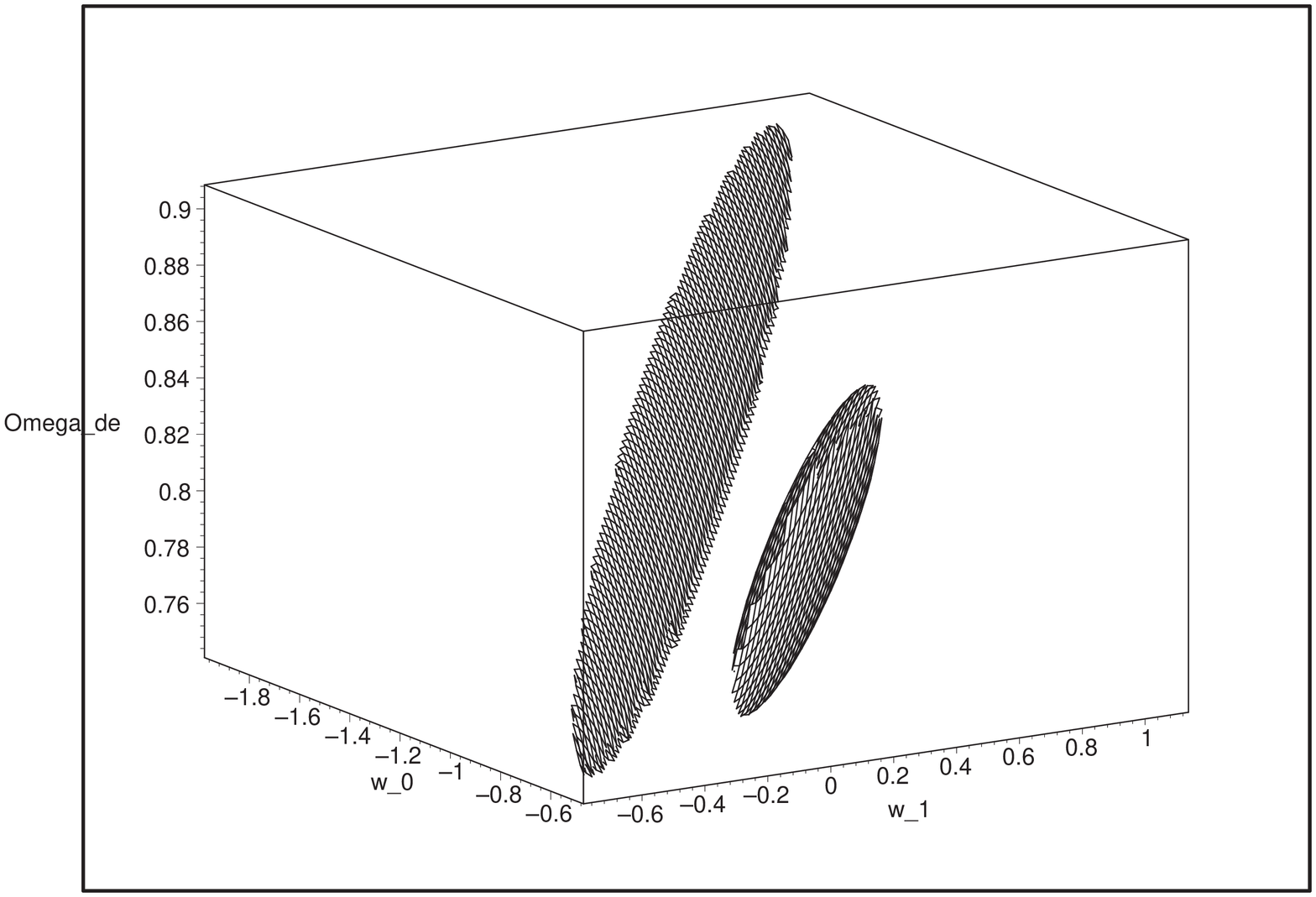}} \\
\end{tabular}
\caption{\label{fig:results} 
LEFT(2D): Best fit equations of state: Solid contours are
    for fits to SN Ia simulated data and the CMB shift parameter $\mathcal{R}$, while 
    dashed contours are for fits to the growth-rate simulated data and the CMB shift parameter.  
RIGHT(3D): Best fit Dark Energy parameter spaces  ($\Omega_{de}$, $w_0$, $w_1$). 
    The ellipsoid to the right of the 3D-figure is for fit to SN Ia simulated data and the CMB 
    shift parameter, while the ellipsoid to the left (3D-figure) is for fits to the growth rate 
    simulated data and the CMB shift parameter data. 
For both figures, the significant difference (inconsistency) between the parameter spaces found using
    the two combinations is due to the DGP model assumed by hypothesis and used to simulate the data. The inconsistency is thus an indication that cosmic acceleration in this case 
    is due to modified gravity at cosmological scales rather then a Dark Energy component. 
}
\end{center}
\end{figure}

Our results, in figure \ref{fig:results},
show that the two allowed regions in the dark energy parameter 
space are significantly different. This signals the expected inconsistency 
between the expansion history and the growth rate of large scale structure. 
The source of the inconsistency is from point \textit{(i)} where the data
was generated using a DGP model, i.e. from our hypothesis that
the true cosmology is that of a modified gravity DGP model. 
Thus, the inconsistency constitute an observational detection of the assumed 
underlying modified gravity model. And finding two significantly 
different equations of state implies that these are not true but forced ones.

The test is based on the comparison of measurements of the expansion history 
and measurements of the growth rate of large scale structure and shows that we can 
go beyond the equation of state analysis. We demonstrated here the working of
consistency tests based on this comparison and provided a preliminary implemention 
in \cite{Ishak2006}. Other works that explored the same idea include \cite{Knox2006,Linder2005}. 
Future work is needed in order to make these tests more 
robust and generic: e.g. to consider other dark energy models (with couplings, 
unusual sound speeds), other modified gravity models, and comparison with systematic 
effects of the probes. Another interesting approach was discussed in reference 
\cite{schimd} where the authors considered signatures of quintessence 
models and their extension to scalar-tensor gravity on weak gravitational lensing
observables.  They found that some models can let an imprint of ten percent on lensing 
observables. The important point from these examples and others is that 
cosmological observations that can probe the growth of cosmological 
perturbations are promising tools to learn about the acceleration of 
the universe beyond the effective equation of state degeneracy.
\subsection{The Weyl-Lovelock theorem as an argument against the full identification of the cosmological constant with vacuum energy} \label{sec:uniqueness}

Based on the following theorems, one could argue against the full identification 
of the cosmological constant with vacuum energy. Indeed, Cartan \cite{cartan}, 
Weyl \cite{weyl} and Vermeil \cite{vermeil} proved different theorems showing that 
the only tensor of valency two, $A_{\alpha \beta}$, that is:

a) constructed from the metric tensor $g_{\alpha \beta}$ and its 
first two partial derivatives, $g_{\alpha \beta,\gamma}$ and 
$g_{\alpha \beta,\gamma \delta}$,

b) divergence free, i.e. $A^{\alpha \beta}\,_{;\beta}=0$,

c) symmetric, i.e. $A_{\alpha \beta}=A_{\beta \alpha}$,

d) linear in the second derivatives of $g_{\alpha \beta}$,

is
\begin{equation}
A_{\alpha \beta}= c_1 G_{\alpha \beta} + c_2 g_{\alpha \beta}
\end{equation}
where $c_1$ and $c_2$  are constants  and $G_{\alpha \beta}$ is the 
Einstein tensor (\ref{eq:einstein}). Lovelock \cite{lovelock1,lovelock2} showed that 
conditions c) and d) are superfluous when the spacetime dimension 
is 4. (Note that when $A_{\alpha \beta}$ is put in the
 EFE, $c_1$ is absorbed in the $\kappa$ factor.)

In Refs. \cite{weyl} and \cite{cartan}, it was first 
proven that the most general curvature tensor $A_{\alpha \beta}$ 
is a linear combination of $R_{\alpha \beta}$, $R g_{\alpha \beta}$ 
and $g_{\alpha \beta}$, i.e. of the form 
\begin{equation}
\label{eq:Aeq}
A_{\alpha \beta}=a R_{\alpha \beta} + b R g_{\alpha \beta} + c
g_{\alpha \beta}, 
\end{equation}
where a, b, and c are constants.
Then values $a=1$ and $b=-\frac{1}{2}$ are 
derived from the divergence free condition (conservation law). 

Consequently, one is tempted to take the standpoint that unless one 
is guided by some physical laws or measurements, setting the constant
 $c$ (i.e. $\Lambda$) in equation (\ref{eq:Aeq}) to any particular value, 
including zero is unjustifiable.

Imposing {\it a priori}  a particular value on $\Lambda$ 
(for instance zero) is perhaps making the same
mistake Einstein did by putting the particular value 
\begin{equation}
\Lambda_{Einstein}=\frac{4}{9 C^2}
\end{equation} 
(for a closed static universe) where $C$ is as defined previously, after equation (\ref{eq:scalefactor}).

Thus, this suggests that a geometrical constant $\Lambda$-term in 
the EFE (\ref{eq:efe}) is part of the equations on its own right 
with no reference to any energy momentum tensor.
This could be used as an argument not in favor of 
the exact identification of the geometrical 
cosmological constant with vacuum energy. 

\subsection{What are the implications of the simplest solution in view of principles of General Relativity? }
As discussed in the previous sections, some of the assumptions
underlying the cosmological constant problems are not 
unquestionable and it is important to find ways to challenge 
them and put them to the test.

Needless to recall,  the simplest solution can arise from, 
first, abandoning the assumed identification 
of the cosmological constant with vacuum energy 
(based on the theorems discussed in the previous sub-section \ref{sec:uniqueness}), 
and second, putting on the side the cosmic coincidence 
(see section \ref{sec:coincidence} for a discussion).
The remaining question is then why the huge vacuum energy densities
(see section \ref{sec:old_problem}) from quantum field theory estimations 
do not contribute to the energy budget in the universe.
This question can be legitimately replaced 
by, how does vacuum energy contribute to the EFE?
For example, is it correct to try to add contributions from vacuum energy density
to the EFE using some ultra-violet cutoff energy?  
We discuss this point more in section \ref{sec:casimir}, see also \cite{pad,milton1,milton2,mazur}.

Further, did we really exhaust all possibility 
of exact cancelation mechanisms for vacuum energy? 
\cite{review1,review2,review3,review4,review5,review6,review7}.

It is perhaps worth mentioning that a rather negative vacuum energy/cosmological 
constant was expected within some candidates for a unified theory such as String 
Theory \cite{Witten}, as there is no attractive way to derive a stable vacuum with 
a positive cosmological constant \cite{Witten}. So, in addition to the magnitude 
problems, could this sign problem be a further indication to revise the full 
identification above?

In this simple solution,  
what is measured currently is simply an intrinsic 
curvature of the spacetime, and the value of $\Lambda$ 
is just a constant measured from experiments, as is 
Newton's gravitational constant $G$. 

At this point, we would like to discuss the following
subtle point. 
The usual aesthetic interpretation of one of General Relativity's principles
is that the mass-energy content of the universe creates curvature of the spacetime 
(assuming a zero Weyl tensor as in standard FLRW cosmology).
One could then ask the following question: if spacetime is to have a curvature 
in absence of mass-energy sources and this curvature is not due to vacuum energy 
then what is generating this curvature? 
There are two possible answers to the question: 
If one wants to preserve the interpretation above, then one
needs to explain the source of this curvature. However, it is 
also correct to take the other standpoint and consider   
that the Einstein field equations are a set of differential 
equations containing a cosmological constant and governing 
the laws of General Relativity, and that 
in absence of sources the trivial spacetime is simply
de Sitter with an intrinsic curvature.

The simple solution we discussed in this section is of the 
latter type and is a consequence of 
questioning some of the assumptions usually made.
To our best knowledge, this particular simple but subtle 
point about loosing the aesthetic interpretation above has 
not been discussed in literature about dark energy.

With these reconsiderations in mind, it is important to think 
about identifying new astrophysical tests or experiments at the 
interface of particle and gravitational physics, as we discuss 
in the next sub-section.

\subsection{Vacuum energy and Casimir effect in gravitational and cosmological contexts}
\label{sec:casimir}
Another possible successful approach to the cosmological constant problems 
is to think of a situation or an experiment where the validity of the cosmological 
constant-vacuum energy identification can be put to the test.  
A geometrical cosmological constant has  no quantum properties while vacuum energy 
has both gravitational and quantum properties. 
Also, is it possible to learn more on how vacuum energy contributes to the 
cosmological constant? Some of these questions started to be addressed in the 
literature as we cite further.

It is perhaps relevant at this point to recall the 
Casimir effect \cite{casimir} which is a purely quantum 
field theory phenomenon (see \cite{casimirreview} for 
a recent comprehensive review, and references therein.) 
The Casimir effect results from a change in the zero-point
oscillations spectrum of a quantized field when the 
quantization domain is restricted or when the topology of 
the space is non-trivial.
For example, a Casimir force appears as the result of the alteration 
of the vacuum energy  by some boundaries. 
In its simplest form, predicted by 
Casimir \cite{casimir}, two neutral plane parallel 
conducting plates placed in a vacuum at a distance $a$
from one another will experience an attractive force 
$F_{_C}=-\frac{\pi^2 \hbar c}{240 a^4}S$ 
where $S \gg a$ is the plate area. The Casimir effect has been now
extensively measured with a few percent precision \cite{casimirreview}.

Cosmologically, the Casimir effect is significant when 
the topology of the model of the universe is non-trivial 
(different from an infinite Euclidean topology), see e.g. \cite{topology1}.
The effect has been discussed in models with non-trivial topology,
notably the simple case of a closed FRW 
universe with a 3-torus topology, see e.g. \cite{torus}. 
Also, the Casimir energy has been used from compact extra
dimensions \cite{milton1,milton2,mazur} to discuss the cosmological 
constant problem, and 
 with models with supersymmetric large extra dimensions
to propose cancelation mechanisms for the cosmological
constant problem, see e.g. \cite{Burgess}.

In a more relevant context for our discussion, 
one would like to study, via the Casimir effect, 
the gravitational properties of the vacuum energy. 
For example, Refs. \cite{caldwell,calloni} calculated correction 
terms to the Casimir force due to the weak gravitational field. 
Such corrections represent the effect of gravitational curvature on 
quantum vacuum fluctuations. The authors of Ref. \cite{calloni} 
evaluated the order of the force acting on a Casimir apparatus 
redshifting in a weak gravitational field and concluded that, 
although some issues with signal modulation need to be solved, 
testing such force should be feasible and within reach of present 
technological resources.

Now, related to our question on how the vacuum energy 
may fit within the EFE, it has been argued in 
some papers, see for example \cite{milton1,milton2,mazur}, that
 as the measured 
Casimir effect is related to vacuum energy differences, 
the vacuum energy may not contribute to the cosmological 
dynamics via some fixed cutoff energy but rather via 
energy differences as in Casimir energy. 
This Casimir 
energy can be produced from some compact extra dimensions \cite{milton1,milton2,mazur} or 
non-trivial topology of the spacetime \cite{topology1}.

We could state that if this is the case, then 
as we have not yet detected any non-trivial topology 
for a wide rang of models \cite{cornish},
 this could imply the vanishing of the vacuum energy
contribution at cosmological scales.
On the other hand, not all non-trivial topologies
have been ruled out and one could push the idea further.

Therefore, questioning and testing how vacuum energy contributes 
to the cosmological constant using, for example, the Casimir effect 
in a cosmological context may prove helpful to the dark energy questions.  
\section{Concluding remarks}

We discussed different formulations of the cosmological constant/dark 
energy problems and some of the assumptions underlying them. We argued 
for the usefulness of clarifying and questioning some of these assumptions 
and identifying new strategies in order to put them to the test. 

We used some fraction calculations in order to evaluate how much of 
a fine-tuning is involved in the cosmic coincidence. We found that these 
fractions are of the order of a few percent even in the worst cases. 
This is significant but not totally unusual in physics and astrophysics. 
Therefore, on one hand, it remains perhaps a motivated and 
interesting problem to seek models that 
could naturally explain these numbers.
On the other hand, it was important 
to clarify that this fine tuning should not constitute a barrier 
that rejects a successful solution for the other 
cosmological constant problems.

Current and future plans are focused on constraining the equation of state 
of dark energy using cosmological probes. This is certainly an important approach 
and some progress has been made, however as we showed in section IV, constraining a 
variable equation of state will require very sophisticated and challanging future 
experiments. Furtheremore, this approach is limited in the kind of decisive 
answers it could provide on the nature of dark energy. Indeed, unless we are lucky 
enough to find a dark energy that has an equation of state significantly different 
from that of a cosmological constant, new kinds of tests or experiments will be 
necessary in order to provide conclusive answers to the dark energy problem.
For example, finding that dark energy parameters are 
those of a cosmological constant to a few percent precision
will be very suggestive but will require tests different from the  
equation of state in order to rule out decisively dynamical 
dark energy models. Now, even if we are ready to accept some 
high level of precision to be satisfactory (or if we reach fundamental 
limitations of our experiments), finding dark energy parameters 
that are characteristic of a cosmological constant will only confirm 
the cosmological constant problems with no further clues. 
Further, once an equation of state is determined from cosmological 
observations, one is always left with the following question: Is this an
effective equation of state of some dark energy component in the 
energy momentum tensor or is this a forced equation of state obtained by 
fitting dark energy models on the top of some modified model of gravity?

Therefore it is important to encourage other directions and strategies for 
approaching the dark energy problems. In particular, we discussed the relevance 
of questioning and challenging some of the assumptions underlying the formulation 
of the cosmological constant problems in order to look for new types of tests.

Next, we showed that comparing cosmological observations of the 
expansion history and cosmological observations of the growth rate of 
large-scale structure can distinguish between cosmic acceleration due to 
some dark energy models and cosmic acceleration due to some modification to 
gravity physics at cosmological scales. The basic idea 
is that the effect of cosmic acceleration on the expansion function and its 
effect on the growth rate function must be mathematically consistent one with 
another because of the underlying gravity theory (General Relativity). As shown in section V-A, the 
failure in the consistency relation can be used as a test to distinguish between 
cosmic acceleration due to dark energy models and acceleration due to modified 
gravity at cosmological scales. This consistency test shows the potential of some 
cosmological probes such as, supernova searches, gravitational lensing and clusters 
of galaxies to go beyond the equation of state approach in order to address the 
cosmological constant/dark energy questions. 

Next, motivated by some theorems on the most general curvature tensor 
in the Einstein field equations, we argued that the identification of 
the cosmological constant with the vacuum energy is not unquestionable 
and might bring some limitations of its own because it changes the formulation 
of the old cosmological constant problem. Recall that as a result of this 
questioning, dark energy can be identified as a simple geometrical cosmological 
constant. In the absence of sources, the trivial spacetime is then de Sitter 
with an intrinsic constant curvature. However, then two questions arise. 
i) An important interpretation of one of principles of General Relativity is that 
the mass-energy content of the universe creates curvature of the spacetime. 
One could then ask the following question: if spacetime is to have a curvature 
in the absence of mass-energy sources, and this curvature is not due to vacuum 
energy, then what is generating this curvature? 
There are two possible answers to this question: If we want to preserve the 
interpretation above then we do need to explain this curvature.
However, it is fully correct to take the other standpoint and consider 
that the Einstein field equations are a set of differential 
equations containing a cosmological constant and governing 
the laws of gravity (General Relativity), and that in the absence 
of sources, spacetime has an intrinsic constant curvature.
This last possibility requires one to sacrifice the important 
interpretation mentioned above.
ii) The second question is why would the huge vacuum energy densities 
evaluated from quantum field theory calculations not contribute 
to the measured effective cosmological constant? As we discussed, 
this question could be re-addressed in the context of how the vacuum 
energy may contribute to the Einstein field equations. In particular,
is the usual method of using a given ultraviolet cutoff energy  
as a source of gravity questionable? For example, other propositions  
have been made in literature \cite{milton1,milton2,mazur} where vacuum 
energy will contribute via energy differences as experienced with 
the Casimir effect.

Finally, we pointed out the possible role of the Casimir effect 
used in gravitational and cosmological contexts for testing 
some of the assumptions and questions discussed.
This is a purely quantum field theory phenomenon and could
be used to look for clues on how the vacuum energy may fit 
within the Einstein field equations. Other kinds of experiments 
at the interface between quantum field theory and general 
relativistic principles have been also discussed in  
\cite{reynaud1,reynaud2,viola,papini} and might be of 
similar interest.

We conclude that challenging some of the 
assumptions underlying the formulation of the cosmological 
constant/dark energy problems and putting them to the test may 
prove useful and necessary to make progress on these questions.

\acknowledgments
The author thanks 
Latham Boyle,
Simon DeDeo,
G.F.R. Ellis, 
Chris Hirata, 
Pat McDonald, 
David Spergel, 
Paul Steinhardt, 
and 
Amol Upadhye for useful comments.
The author thanks James Richardson for reading the manuscript. 

This is not a review paper and we acknowledge that the list of references cited here is incomplete. We tried to provide only some examples from the literature when necessary. 

Partial support from the Natural Sciences and Engineering Research 
Council of Canada (NSERC) and NASA Theory Award NNG04GK55G at Princeton University
is acknowledged. The author acknowledges the partial support from the Hoblitzelle Foundation and a Clark award at the University of Texas at Dallas.

\appendix 
\section{Examples of spacetime curvature including a cosmological constant}
\label{app:curvature}
\begin{figure}
\begin{center}
\begin{tabular}{cc}
{\includegraphics[width=2.3in,angle=-90]{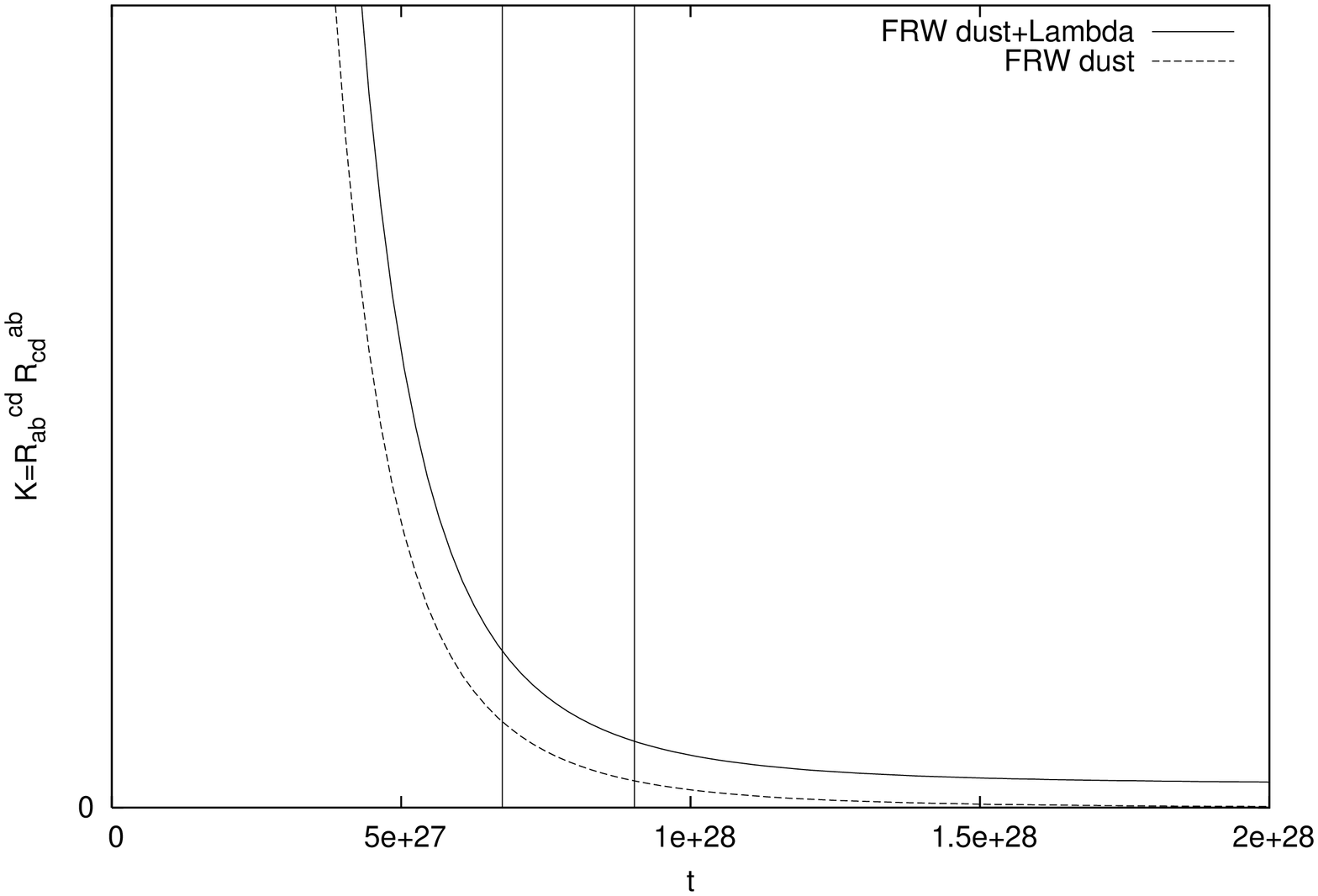}} &
{\includegraphics[width=2.3in,angle=-90]{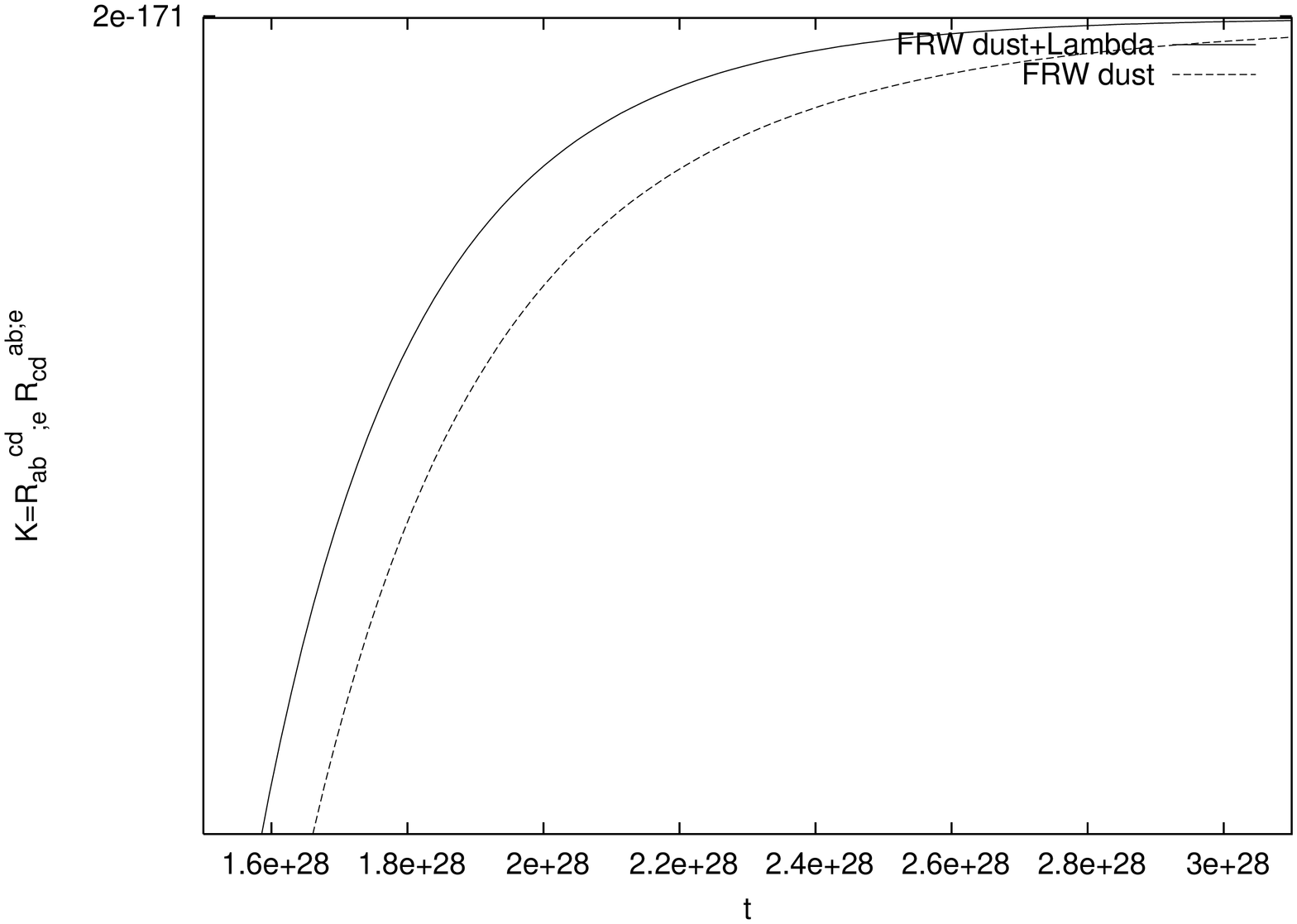}} \\
\end{tabular}
\caption{\label{fig:frwdesitterRiemSqAB} 
a) Plot of the curvature invariant $\mathcal{K}=R_{\alpha \beta}\,^{\gamma \delta}\,\;R_{\gamma \delta}\,^{\alpha \beta}$.
b) Plot of the differential invariant $DiRiem=R_{\alpha \beta}\,^{\gamma \delta}\,_{ ; \eta}\;R_{\gamma \delta}\,^{\alpha \beta}\,^{;\eta}$. 
\\
The curvature decreases during a matter dominated universe to 
reach a constant curvature Lambda-dominated universe. Length 
units are used with $\Lambda=10^{-56}cm^{-2}$. We also display on the left vertical lines for $ct_{transition}=0.67\times 10^{28}cm$ and $ct_{\rho_{\Lambda}=\rho_m}=0.90\times 10^{28}cm.$
} 
\end{center}
\end{figure}
In order to plot the evolution of spacetime curvature with a cosmological constant, 
we consider curvature invariants constructed from the Riemann tensor,
$R_{\alpha \beta \gamma \delta}$.
These scalars 
allow a coordinate independent study 
of some geometrical features of a spacetime. 
They can also be linked to physical quantities via 
the EFE.
For the special spacetimes we consider here 
the invariants are all related via algebraic relations \cite{carot,santo},
and for the sake of simplicity we just choose here the Kretchman scalar,  
\begin{equation}
\label{eq:K}
\mathcal{K}=R_{\alpha \beta}\,^{\gamma \delta}\,\;R_{\gamma \delta}\,^{\alpha \beta}
\end{equation}
and the differential invariant
\begin{equation}
\label{eq:DiRiem}
DiRiem=R_{\alpha \beta}\,^{\gamma \delta}\,_{ ; \eta}\;R_{\gamma \delta}\,^{\alpha \beta}\,^{;\eta}
\end{equation}
to trace the evolution of the curvature of spacetime. For the metric (\ref{eq:frwdesitter}), the invariants read
\begin{equation}
\label{eq:frwdesitterRiemSq}
\mathcal{K}=\Lambda^2\Big{[}\frac{5}{3}\Big(\coth\big{(}\frac{\sqrt{3\Lambda}}{2}t\big{)}\Big)^4-2\Big(\coth\big{(}\frac{\sqrt{3\Lambda}}{2}t\big{)}\Big)^2+3\Big{]}
\end{equation}
and 
\begin{equation}
\label{eq:frwdesitterDiRiem}
DiRiem=-9\Lambda^3 \frac{\Big(\coth\big{(}\frac{\sqrt{3\Lambda}}{2}t\big{)}\Big)^2 }
{\Big(\sinh\big{(}\frac{\sqrt{3\Lambda}}{2}t\big{)}\Big)^4}.
\end{equation}
For the matter dominated universe, these are simply given by, 
 $\mathcal{K}=\frac{80}{37}\frac{1}{t^4}$ and $DiRiem=-\frac{60}{3}\frac{1}{t^6}$.
The de Sitter space has the usual line element
\begin{equation}
\label{eq:desitter}
ds^2=-dt^2+\exp{\big(2\sqrt{\frac{\Lambda}{3}}t\big)}(dr^2+r^2 d\Omega^2)
\end{equation}
and constant curvature with $\mathcal{K}=\frac{3}{8}\Lambda^2$.

Figure \ref{fig:frwdesitterRiemSqAB}a and Figure \ref{fig:frwdesitterRiemSqAB}b 
show the profile of $\mathcal{K}$ and $DiRiem$ and how the 
spacetime curvature decreases during the expanding matter dominated universe 
to reach a constant curvature $\Lambda$-dominated universe at late times. 
This can also be seen from taking the limits of 
(\ref{eq:frwdesitterRiemSq}) and (\ref{eq:frwdesitterDiRiem}) at very large $t$.

The vertical lines in Figure \ref{fig:frwdesitterRiemSqAB}a are 
the time at equality of dark energy density with matter energy density 
and the time of transition from deceleration to acceleration.
The no-Lambda curves are shown for comparison.

Furthermore, in order to trace some features of the curvature lost 
in the squared quantities, 
we recourse to plotting directly the non-vanishing components
of the Riemann tensor. Though coordinate dependent, these can 
be informative \cite{book1}. 
\begin{figure}
\begin{center}
\begin{tabular}{cc}
{\includegraphics[width=2.3in,angle=-90]{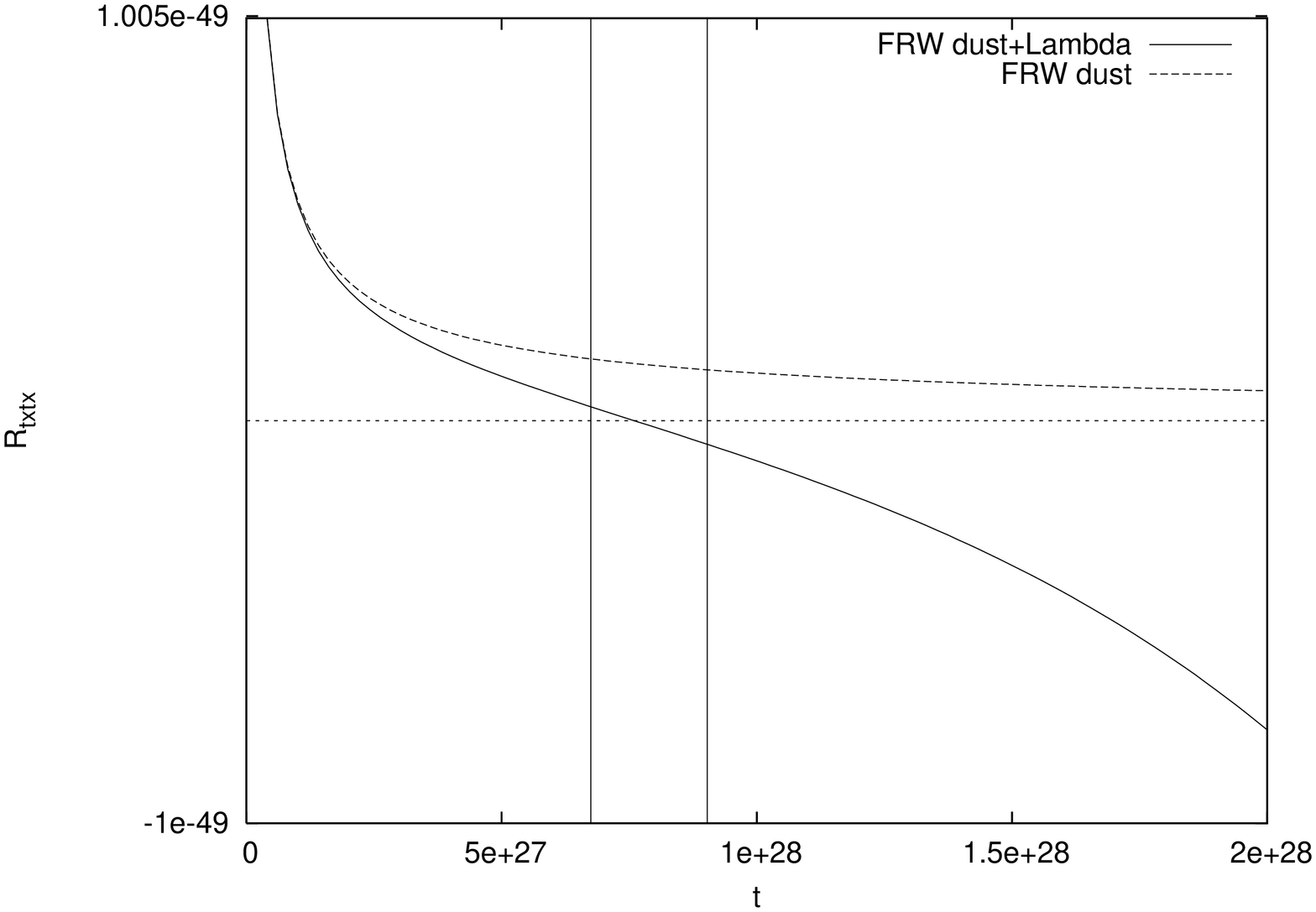}} &
{\includegraphics[width=2.3in,angle=-90]{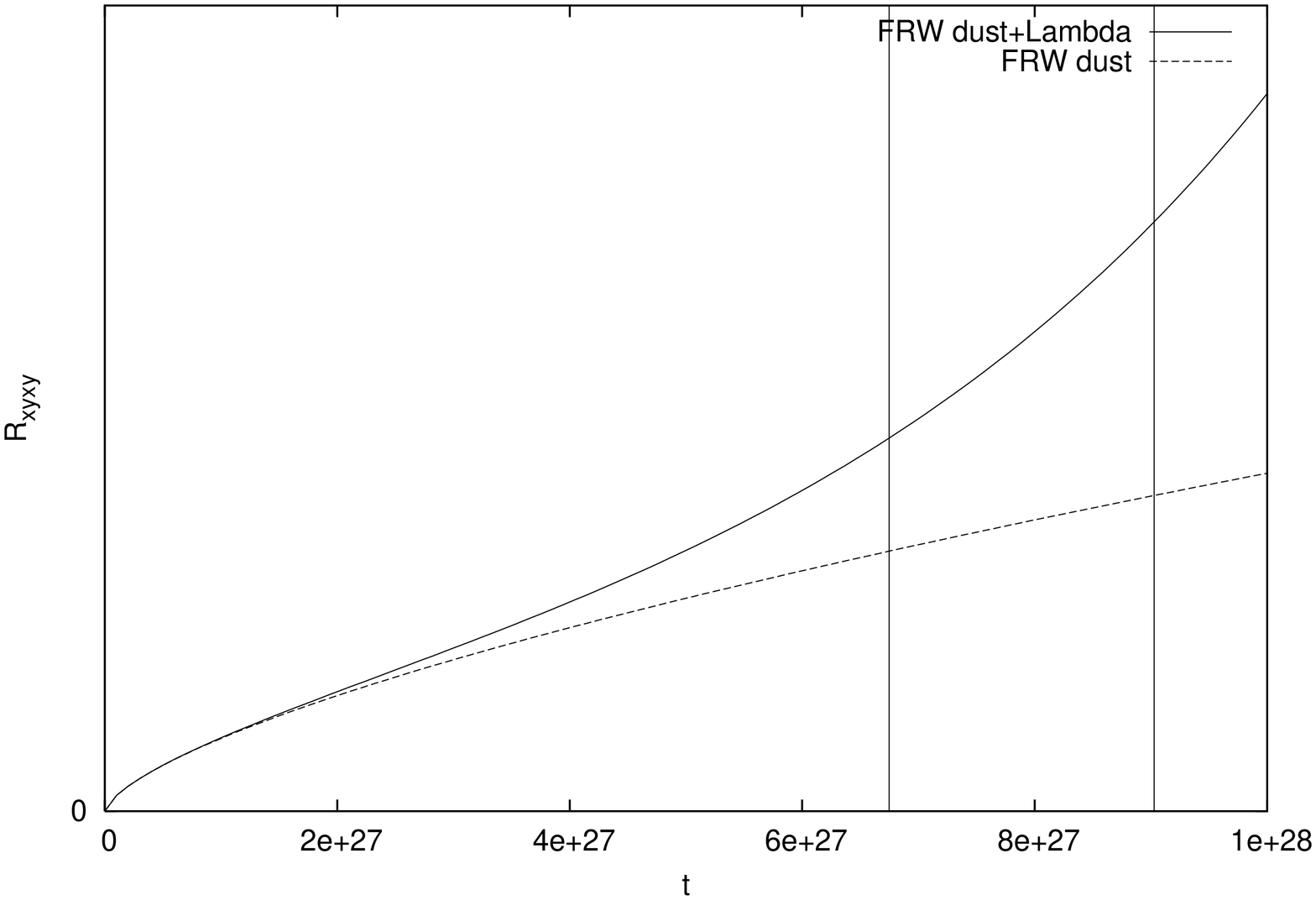}} \\
\end{tabular}
\caption{\label{fig:frwdesitterRiemann12} 
a) LEFT: Plot of the Riemnann components $R_{txtx}=R_{tyty}=R_{tztz}$.
It starts with a
power law decrease to reach a negative range exponential 
decrease during a de Sitter phase. For comparison, the 
no-Lambda curve shows how this component continues with
a power law decrease within a positive range.
Length units are used with $\Lambda=10^{-56}cm^{-2}$. We display vertical for
$ct_{transition}=0.67\times 10^{28}cm$ and $ct_{\rho_{\Lambda}=\rho_m}=0.90\times 10^{28}cm.$ 
\\
b) RIGHT: Plot of the Riemann components $R_{xyxy}=R_{xzxz}=R_{yzyz}$.
The profile is similar to that of the 
 scale factor. These components transit to an exponential 
increase at very large t.
 We also display
$ct_{transition}=0.67\times 10^{28}cm$ and $ct_{\rho_{\Lambda}=\rho_m}=0.90\times 10^{28}cm.$
} 
\end{center}
\end{figure}

In cartesian coordinates, these are:
\begin{eqnarray}
R_{txtx}=R_{tyty}=R_{tztz}=\frac{\Lambda}{6}\Big(\frac{3C}{\Lambda}\Big)^{2/3} \times \\ \nonumber
\Big[ \Big(\sinh\big{(}\frac{\sqrt{3\Lambda}}{2}t\big{)}\Big{)}^{-2/3}-2
\Big(\sinh\big{(}\frac{\sqrt{3\Lambda}}{2}t\big{)}\Big{)}^{4/3}\Big]
\end{eqnarray}
and 
\begin{eqnarray}
R_{xyxy}=R_{xzxz}=R_{yzyz}=3^{1/3}\Lambda\Big(\frac{C}{\Lambda}\Big)^{4/3} \times \\ \nonumber
\Big[ \Big(\sinh\big{(}\frac{\sqrt{3\Lambda}}{2}t\big{)}\Big{)}^{8/3}+
\Big(\sinh\big{(}\frac{\sqrt{3\Lambda}}{2}t\big{)}\Big{)}^{2/3}\Big]
\end{eqnarray}
The time evolution of the $R_{txtx}$ components is shown in 
Figure \ref{fig:frwdesitterRiemann12}a where it starts with a
power law decrease to reach a negative range exponential 
decrease during a de Sitter phase. For comparison, the 
no-$\lambda$ curve shows how this component continues with
 power law decrease within a positive range.

Figure \ref{fig:frwdesitterRiemann12}b shows that the profile
of the $R_{xyxy}$ component is similar to that of the 
 scale factor. This component transits to an exponential 
increase at very large t.

Also, we consider another informative example, 
the Schwarzschild-de Sitter spacetime with
\begin{eqnarray}
\label{eq:schwdesitter}
ds^2=-\Big(1-\frac{2m}{r}-\frac{\Lambda r^2}{3}\big) dt^2+
\\  \nonumber \; \; \; \; \; 
\Big(1-\frac{2m}{r}-\frac{\Lambda r^2}{3}\big)^{-1} dr^2+r^2 d\Omega^2.
\end{eqnarray}
\begin{figure}
\includegraphics[width=2.3in,angle=-90]{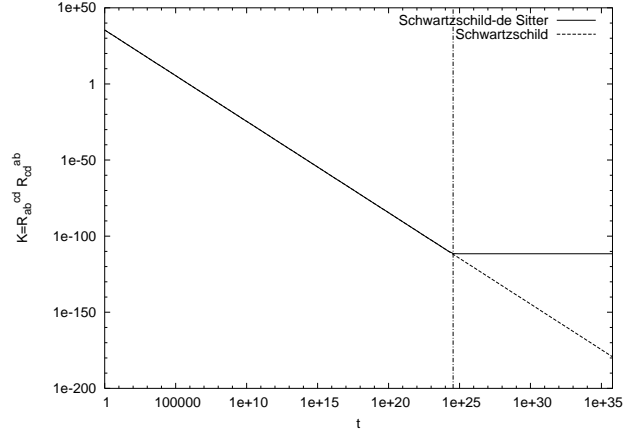}
\caption{\label{fig:schwdesitterRiemSq} 
Plot of $\mathcal{K}=R_{\alpha \beta}\,^{\gamma \delta}
\,\;R_{\gamma \delta}\,^{\alpha \beta}$ as a function of $r$.
The curvature decreases as $1/r^2$ from the 
central mass $m$ to become dominated by the $\Lambda$
term after $r_{|\frac{8}{3}\Lambda^2|=|\frac{48m^2}{r^2}|}=.36\times 10^{25}cm$. 
Length units are used with 
$\Lambda=10^{-56}cm^{-2}$ and a mass $m=0.74\times 10^{17}cm$. 
 }
\end{figure}
At large $r$, it tends to the de Sitter space limit.
The explicit de Sitter case is obtained by setting $m=0$ while the 
explicit Schwarzschild case is obtained by setting $\Lambda=0$.
Here, for (\ref{eq:schwdesitter}) 
\begin{equation}
\mathcal{K}=\frac{8}{3}\Lambda^2+\frac{48m^2}{r^2}
\end{equation}
and is plotted in Figure (\ref{fig:schwdesitterRiemSq})  
where the curvature due to the central mass $m$ 
decreases as a function of r and is overtaken by the 
$\Lambda$ term at very large $r$. The Schwarzschild
curve is plotted for comparison. 
{}
\end{document}